\documentclass[]{aa}  

\usepackage{graphicx}
\usepackage{txfonts}
\newcommand{\angstrom}{\mbox{\normalfont\AA}}
\usepackage{amsmath}	
\usepackage{amssymb}	
\usepackage[colorlinks,citecolor=blue]{hyperref}
\usepackage{graphicx}	
\usepackage{multirow}  
\usepackage{xcolor}

\begin{document}

\title{Transient obscuration event captured in NGC\,3227}

\subtitle{III. Photoionization modeling of the X-ray obscuration event in 2019}

\author{Junjie Mao\inst{\ref{inst:hu0},\ref{inst:strath},\ref{inst:sron}} 
J.~S. Kaastra \inst{\ref{inst:sron},\ref{inst:leiden}},
M. Mehdipour \inst{\ref{inst:stis}}, 
G. A. Kriss \inst{\ref{inst:stis}}, 
Yijun Wang \inst{\ref{inst:ustc0}, \ref{inst:ustc1}, \ref{inst:sron}, \ref{inst:leiden}, \ref{inst:nju0}, \ref{inst:nju1}}, 
S. Grafton-Waters \inst{\ref{inst:mssl}}, \\
G. Branduardi-Raymont \inst{\ref{inst:mssl}}, 
C. Pinto \inst{\ref{inst:inaf0}}, 
H. Landt \inst{\ref{inst:durham}}, 
D.~J. Walton \inst{\ref{inst:ioa}},
E. Costantini \inst{\ref{inst:sron}}, 
L. Di Gesu \inst{\ref{inst:asi}}, 
S. Bianchi \inst{\ref{inst:roma3}}, \\
P.-O. Petrucci \inst{\ref{inst:cnrs}},
B. De Marco \inst{\ref{inst:upc}}, 
G. Ponti \inst{\ref{inst:inaf1}, \ref{inst:mpe}},
Yasushi Fukazawa \inst{\ref{inst:hu0}},
J. Ebrero \inst{\ref{inst:esac}},
and E. Behar \inst{\ref{inst:tiit}}
}

\institute{Department of physics, Hiroshima University, 1-3-1 Kagamiyama, Higashi-Hiroshima, Hiroshima 739-8526, Japan. \label{inst:hu0} \\
\email{jmao2018@hiroshima-u.ac.jp}
\and
Department of Physics, University of Strathclyde, Glasgow G4 0NG, UK \label{inst:strath}
\and
SRON Netherlands Institute for Space Research, Niels Bohrweg 4, 2333 CA Leiden, the Netherlands \label{inst:sron}
\and Leiden Observatory, Leiden University, PO Box 9513, 2300 Leiden, The Netherlands \label{inst:leiden}
\and Space Telescope Science Institute, 3700 San Martin Drive, Baltimore, MD 21218, USA \label{inst:stis}
\and CAS Key Laboratory for Research in Galaxies and Cosmology, Department of Astronomy, University of Science and Technology of China, Hefei 230026, China \label{inst:ustc0}
\and School of Astronomy and Space Science, University of Science and Technology of China, Hefei 230026, China \label{inst:ustc1}
\and Department of Astronomy, Nanjing University, Nanjing 210093, China \label{inst:nju0}
\and Key Laboratory of Modern Astronomy and Astrophysics (Nanjing University), Ministry of Education, Nanjing 210093, China \label{inst:nju1}
\and Mullard Space Science Laboratory, University College London, Holmbury St. Mary, Dorking, Surrey RH5 6NT, UK \label{inst:mssl}
\and INAF-IASF Palermo, Via U. La Malfa 153, I-90146 Palermo, Italy \label{inst:inaf0}
\and Centre for Extragalactic Astronomy, Department of Physics, Durham University, South Road, Durham DH1 3LE, UK \label{inst:durham}
\and Institute of Astronomy, University of Cambridge, Madingley Road, Cambridge CB3 0HA, UK \label{inst:ioa}
\and Italian Space Agency (ASI), Via del Politecnico snc, 00133, Roma, Italy \label{inst:asi}
\and Dipartimento di Matematica e Fisica, Università degli Studi Roma Tre, via della Vasca Navale 84, 00146 Roma, Italy \label{inst:roma3}
\and Univ. Grenoble Alpes, CNRS, IPAG, 38000 Grenoble, France \label{inst:cnrs}
\and Departament de F\'{i}sica, EEBE, Universitat Polit\`{e}cnica de Catalunya, Av. Eduard Maristany 16, E-08019 Barcelona, Spain \label{inst:upc}
\and INAF-Osservatorio Astronomico di Brera, Via E. Bianchi 46, I-23807 Merate (LC), Italy \label{inst:inaf1}
\and Max Planck Institute fur Extraterrestriche Physik, 85748, Garching, Germany \label{inst:mpe}
\and Telespazio UK for the European Space Agency (ESA), European Space Astronomy Centre (ESAC), Camino Bajo del Castillo, s/n, 28692 Villanueva de la Ca\~{n}ada, Madrid, Spain \label{inst:esac}
\and Department of Physics, Technion-Israel Institute of Technology, 32000 Haifa, Israel \label{inst:tiit}
}

\date{Received September 15, 1996; accepted March 16, 1997}

 
\abstract
{A growing number of transient X-ray obscuration events in type I active galactic nuclei (AGN) suggest that our line-of-sight to the central engine is not always free. Multiple X-ray obscuration events have been reported in the nearby Seyfert 1.5 galaxy NGC\,3227 from 2000 to 2016. In late 2019, another X-ray obscuration event was identified with \textit{Swift}. Two coordinated target-of-opportunity observations with XMM-\textit{Newton}, \textit{NuSTAR}, and the Hubble Space Telescope (HST) Cosmic Origins Spectrograph (COS) were triggered in November and December 2019 to study this obscuration event. }
{We aim to constrain the physical properties of the absorbing material (namely, obscurer) that causes the X-ray obscuration event in late 2019. We also aim to compare the handful of obscuration events in NGC\,3227 and other Seyfert galaxies. }
{For each observation, we analyze the time-averaged X-ray spectra collected with XMM-\textit{Newton} and \textit{NuSTAR}. We perform photoionization modeling with the SPEX code, which allows us to constrain the intrinsic continuum simultaneously with various photoionized absorption and emission components. }
{Similar to previous transient X-ray obscuration events in NGC\,3227, the one caught in late 2019 is short-lived (less than five months). If the obscurer has only one photoionized component, the two X-ray observations in late 2019 cannot be explained by the same obscurer that responds to the varying ionizing continuum. Due to the unknown geometry of the obscurer, its number density and distance to the black hole cannot be well constrained. The inferred distance covers at least two orders of magnitude, from the broad-line region (BLR) to the dusty torus. Unlike some other X-ray obscuration events in Seyfert galaxies like NGC\,5548 and NGC\,3783, no prominent blueshifted broad absorption troughs were found in the 2019 HST/COS spectra of NGC\,3227 when compared with archival UV spectra. This might be explained if the X-ray obscurer does not intercept our line of sight to (a significant portion of) the UV emitting region. It is not straightforward to understand the variety of the observational differences of the X-ray obscuration events observed so far. Future observations with high-quality data are needed to unveil the nature of the X-ray obscuration events. }
{}
\keywords{X-rays: galaxies -- galaxies: active -- galaxies: Seyfert -- galaxies: individual: \object{NGC\,3227} -- techniques: spectroscopic}

\titlerunning{Transient X-ray obscuration event of NGC\,3227 in 2019}
\authorrunning{J. Mao et al.}       
\maketitle
%

\section{Introduction}
AGN are the energetic power houses at the centers of active galaxies \citep{net15}. Matter can flow towards the supermassive black hole via accretion. At the same time, matter can also flow outward from the black hole \citep{cre03,lah21}, regulating the accretion process as well as providing feedback to the host galaxy. With detectable emission across the entire electromagnetic spectrum, AGN have many interesting properties and thus many subclasses \citep{pad17}. One of the main classifiers is the dusty torus \citep{ant93}. Type I AGN are those whose line-of-sight towards the central engine is not blocked by the dusty torus. 

A growing number of transient X-ray obscuration events have been reported in Type I AGN \citep[e.g.,][]{lam03,ris07,ris11,mar14,kaa14,ebr16,meh17,lon19,gal21,mil21,ser21}. These events indicate that, in at least some type I AGN, our line-of-sight to the central engine is not always free. During the transient X-ray obscuration events, one of the key features is the significant lowering of the soft X-ray continuum. Narrow emission lines in the soft X-ray band, previously hidden under the relatively high continuum level, might become observable. Furthermore, such X-ray obscuration events might be accompanied by the emergence and variations of absorption features in the UV  \citep[e.g.,][]{kaa14,ebr16,meh17,lon19,kar21,sae21} and NIR spectra \citep{lan19,wil21}.

NGC\,3227 is a nearby \citep[$z=0.003859$,][]{dva91} Seyfert 1.5 galaxy, which contains a supermassive black hole \citep[$M_{\rm BH}=5.96\times10^6~M_{\odot}$,][]{ben15}. Multiple X-ray obscuration events have been reported in NGC\,3227 from 2003 to 2016 \citep{lam03,mar14,beu15,tur18}. As described in \citet[][Paper I hereafter]{meh21}, a \textit{Swift} \citep{geh04} monitoring campaign was carried out in XMM-\textit{Newton} Cycle 18 (PI: J. S. Kaastra) to catch transient X-ray obscuration events in a sample of type I AGN. In late 2019, an X-ray obscuration event in NGC\,3227 was identified with the \textit{Swift} monitoring. We triggered two joint target-of-opportunity observations with XMM-\textit{Newton} \citep{jan01}, \textit{NuSTAR} \citep{har13}, and HST Cosmic Origins Spectrograph \citep[COS,][]{gre12} on 2019-11-15 and 2019-12-05, respectively. 

Taking advantage of the multi-wavelength data of NGC\,3227 obtained in late 2019, we constructed the broadband spectral energy distribution of NGC\,3227 in Paper I. In \citet[][Paper II hereafter]{wang22}, we analyzed the archival XMM-\textit{Newton} data of NGC\,3227 to characterise the warm absorber observed in the X-ray band in the absence of the obscuration event. In Paper II, we also describe transient obscuration events in December 2006 and in late 2016. The former was not identified in studies prior to Paper II, while the latter was reported by \citet{tur18}. In the present work, we analyzed time-averaged X-ray spectra for each observation in November and December 2019 to study the physical properties of the absorbing material (namely, obscurer) that caused the X-ray obscuration event. Variability of the obscurer, as well as the continuum, within each observation will be studied by Grafton-Waters et al. in prep. (Paper IV hereafter). Detailed analysis of the HST/COS spectra will be presented in a future paper by our team. 

\section{Observation}
\label{sct:data_obs}
\begin{figure*}[!h]
\centering
\includegraphics[width=.7\hsize, trim={0.8cm 0.5cm 1.5cm 1.0cm}, clip]{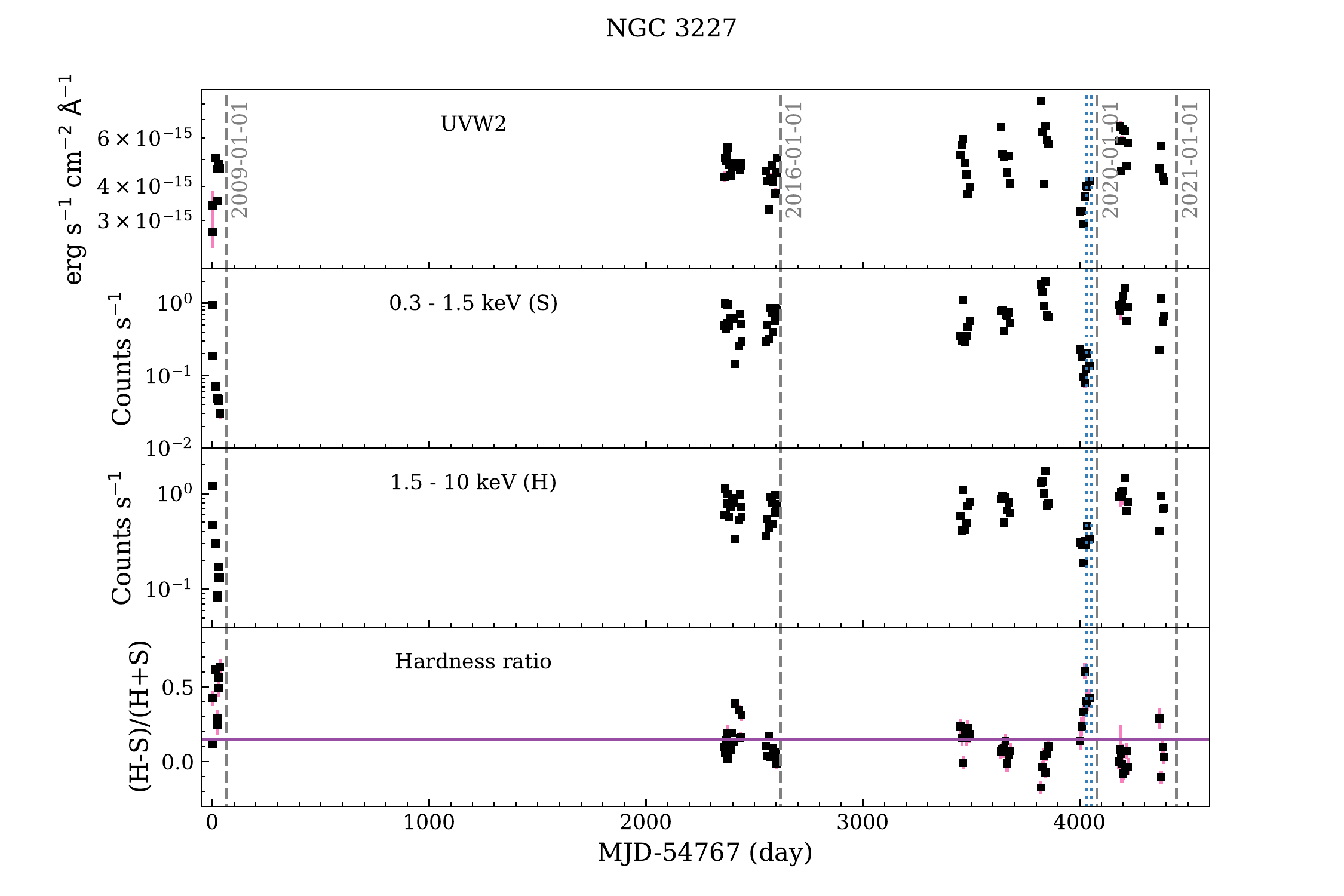}
\caption{Archival \textit{Swift} data for NGC\,3227 from 2008-10-28 to 2020-12-27. The top panel is the UVW2 flux. The two middle panels are the count rates in the hard (H: $1.5-10$~keV) and soft (S: $0.3-1.5$~keV) X-ray bands. Statistical uncertainties are in general too small to be visible in the plot. The bottom panel shows the X-ray hardness ratio $(H-S)/(H+S)$. The horizontal solid line in purple is the historical average hardness ratio before late 2020. Calendar dates are marked by the vertical dashed lines. The vertical dotted lines in blue mark the joint HST/COS, XMM-\textit{Newton}, and \textit{NuSTAR} observations in late 2019.} 
\label{fig:plot_ltc_uvx}
\end{figure*}

\begin{figure*}[!h]
\centering
\includegraphics[width=.7\hsize, trim={0.5cm 0.5cm 1.8cm 1.0cm}, clip]{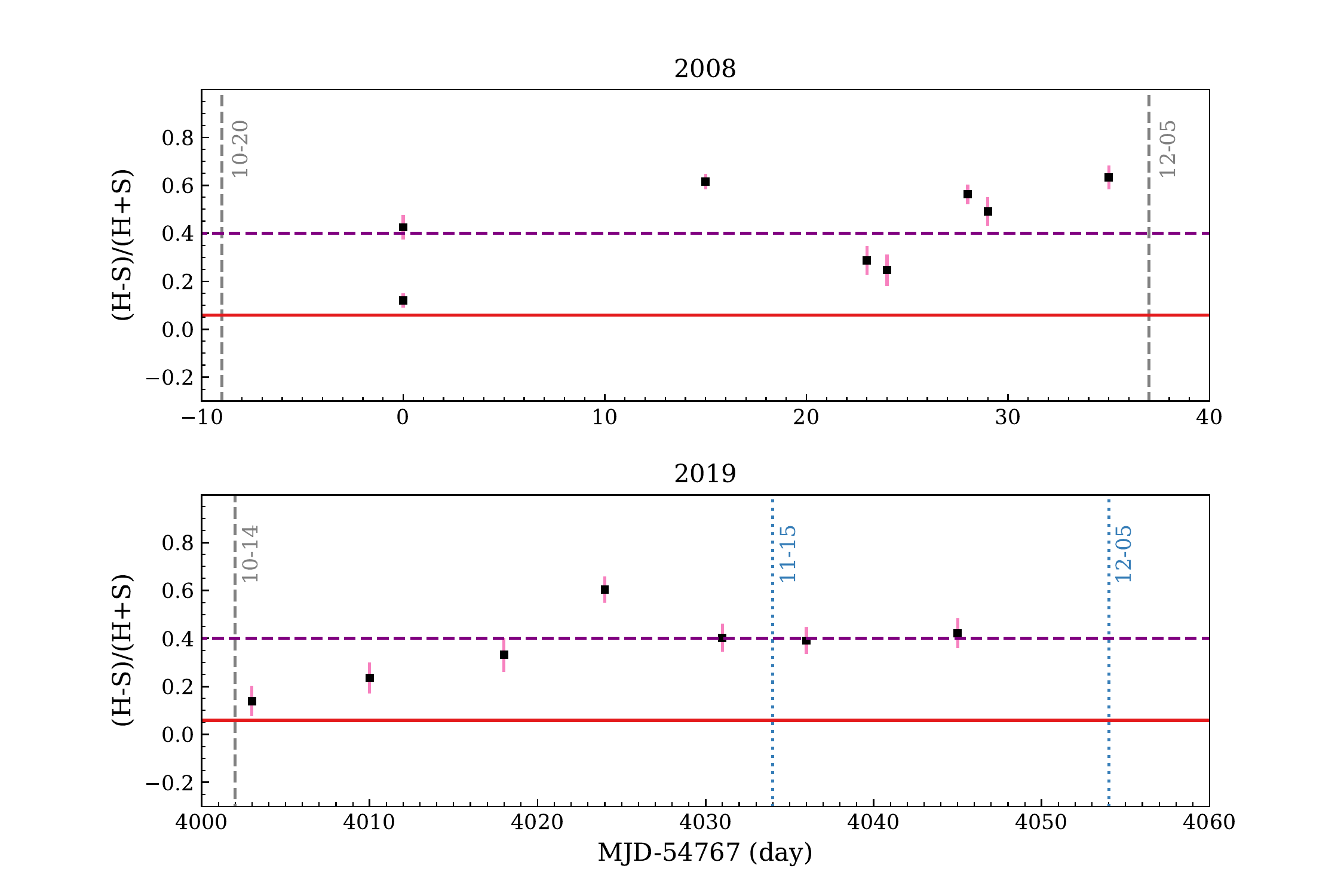}
\caption{\textit{Swift} X-ray hardness ratio $(H-S)/(H+S)$ for NGC\,3227 in 2008 (top) and 2019 (bottom). The horizontal solid line in purple is the historical average hardness ratio before late 2020. The horizontal dashed line in purple, corresponding to $(H-S)/(H+S)=0.4$, is shown merely to guide the eye. Calendar dates are marked by the vertical dashed lines. The vertical dotted lines in blue mark the joint HST/COS, XMM-\textit{Newton}, and \textit{NuSTAR} observations in late 2019.} 
\label{fig:plot_hr_08vs19}
\end{figure*}

In Fig.~\ref{fig:plot_ltc_uvx}, we show the archival Swift data for NGC\,3227 from 2008-10-28 to 2020-12-27. During X-ray obscuration events, the significantly lowered soft X-ray flux will lead to an elevated X-ray hardness ratio. The X-ray hardness ratio of NGC\,3227 increased by $\sim0.5$ in $\sim2$ weeks around early November 2019 (Fig.~\ref{fig:plot_hr_08vs19}, bottom panel). It remained at a high-level until December 5, 2019. After that, the target was out of the Swift visibility window until April 2020, where the hardness ratio was negative. Accordingly, the duration of the X-ray obscuration event in 2019 was less than five months. Note that the timescales of both the elevation and duration can be even shorter for NGC\,3227. Around November 2008, the Swift X-ray hardness ratio increased by $\sim0.3$ within a day and $\sim0.5$ in $\sim2$ weeks (Fig.~\ref{fig:plot_hr_08vs19}, upper panel). On 2016-12-09, the duration of the X-ray obscuration was $\sim20$~ks (Paper II). 

The relatively large Swift X-ray hardness ratio in late 2019 triggered two coordinated multi-wavelength observations with HST/COS, XMM-\textit{Newton}, and \textit{NuSTAR} of NGC\,3227. Table~\ref{tbl:obs_log} lists the XMM-\textit{Newton} and \textit{NuSTAR} data used in the present work. The data reduction was the same as described in Section 2 of Paper I. In Table~\ref{tbl:obs_log}, we also list four HST observations, one in 2000, one in 2010 and two in late 2019. With these UV grating spectra, we aim to briefly investigate the X-ray and UV connection of the obscuring event (Sect.~\ref{sct:dis}). 


\begin{table*}
\caption[]{Observation log.}
\label{tbl:obs_log}
\centering
\begin{tabular}{ccccccc}
\hline\hline
\noalign{\smallskip}
Observatory & Obs. ID & Date & Duration \\
\noalign{\smallskip}
\hline
\noalign{\smallskip}
HST/COS & LDYC02 & 2019-12-05 & 3.7~ks \\
\noalign{\smallskip}
XMM-Newton & 0844341401 & 2019-12-05 & 52~ks  \\
\noalign{\smallskip}
NuSTAR & 80502609004 & 2019-12-05 & 28~ks \\
\noalign{\smallskip}
\hline
\noalign{\smallskip}
HST/COS & LDYC01 & 2019-11-15 & 3.7~ks \\
XMM-Newton & 0844341301 & 2019-11-15 & 105~ks  \\
\noalign{\smallskip}
NuSTAR & 80502609002 & 2019-11-17 & 29~ks \\
\noalign{\smallskip}
\hline
\noalign{\smallskip} 
HST/COS & LB9N01 & 2010-05-10 & 4.4~ks \\ 
\noalign{\smallskip} 
HST/STIS & O5KP01 & 2000-02-08 & 3.9~ks \\ 
\noalign{\smallskip}
\hline
\end{tabular}
\end{table*}

For each observation, the two first-order spectra of Reflection Grating Spectrometer \citep[RGS,][]{dhe01} were fitted simultaneously over the $6-37$~\AA\ wavelength range. The EPIC/pn spectrum in the $1.8-10$~keV energy range was used. The NuSTAR spectra from the two detector modules (FPMA and FPMB) are combined and fitted over the $5-78$~keV energy range. To correct for the cross calibration of different instruments, the following scaling parameters were used. These scaling parameters were obtained by matching the flux level in common energy ranges less affected by emission and absorption features: $8-10$~\AA\ to match RGS and EPIC/pn and $7-10$~keV to match EPIC/pn and NuSTAR. For 2019-11-15, the scaling parameters were 1.0 (RGS1), 1.0 (RGS2), 1.038 (EPIC/pn), and 1.027 (NuSTAR), respectively. For 2019-12-05, the scaling parameters were 1.00 (RGS1), 1.041 (RGS2), 1.062 (EPIC/pn), and 1.073 (NuSTAR), respectively. HST/COS spectra are not included in our spectral analysis. 

\section{X-ray spectral analysis}
\label{sct:spec}
We used SPEX v3.05.00 \citep{kaa18} and $C$-statistics for the X-ray spectral analysis \citep{kaa17}. One of the key features of the photoionization modeling with the SPEX code is to constrain the intrinsic continuum simultaneously with the absorption (and obscuration) effects. Paper I described the baseline model, which we briefly summarize here. The intrinsic broadband spectral energy distribution (SED) of NGC\,3227 consists of a disk blackbody component (\textit{dbb}) that dominates the optical to UV band, a warm Comptonized disk component (\textit{comt}) for the soft X-ray excess, a power-law component (\textit{pow}), and a neutral reflection component (\textit{refl}) for the hard X-ray band. The intrinsic continuum is absorbed by the obscurer, warm absorber, and the Galactic absorption. Both the obscurer and warm absorber are assumed to be photoionized and modelled with \textit{pion} \citep{meh16b} components. The Galactic absorption (by neutral gas) was modeled with a \textit{hot} component with its temperature and hydrogen column density frozen to $0.5$~eV and $N_{\rm H}=2.07\times10^{20}~{\rm cm^{-2}}$ \citep{mur96}, respectively. The protosolar abundances of \citet{lod09} are used for all the plasma models. 

In NGC 5548 \citep{kaa14,whe15,mao18}, NGC 3783 \citep{meh17,mao19}, and Mrk 335 \citep{lon19,par19}, due to the presence of the obscurer, emission features of the warm emitter \citep{tur96} stand out above the reduced soft X-ray continuum. The physical origin of the warm emitter is not clearly understood though. Note that the warm emitter is not included in Paper I, which used the EPIC/pn spectra for the soft X-ray band data. Narrow emission lines in the soft X-ray band are not resolved by EPIC/pn but its large effective area is useful when building the broadband SED model. The RGS spectra used in the present work to resolve emission line features from the warm emitter. Accordingly, we included an emission \textit{pion} component \citep{mao18} for the warm emitter. Due to the relatively low signal to noise ratio of the line features, we reduced the free parameters of the emission \textit{pion} component. The emission covering factor $C_{\rm em}=\Omega/4\pi$ depends on the solid angle ($\Omega$) subtended by the warm emitter with respect to the central engine. We assume a fiducial value of $C_{\rm em}=0.01$, which is within the range of $10^{-4}-10^{-1}$ \citep[e.g.,][]{mao18,mao19,gwa20}. Default values of $v_{\rm out}=0~{\rm km~s^{-1}}$ and $v_{\rm mic}=100~{\rm km~s^{-1}}$ were used\footnote{In SPEX v3.05, the microscopic turbulent velocity $v_{\rm mic}=\sqrt{2}\sigma_{\rm turb}$, where $\sigma_{\rm turb}$ is the root-mean square (RMS) of the line-of-sight velocity.}. The best-fit parameters ($N_{\rm H}$, $\log \xi$\footnote{Throughout this work, the ionization parameter \citep{tar69,kro81} is defined as $\xi = L/(n_{\rm H}r^2)$, where $L$ is the $1-1000$~Ryd ionizing luminosity, $n_{\rm H}$ the hydrogen number density, and $r$ the distance to the black hole.}, and broadening velocity due $v_{\rm mac}$ to macroscopic motion around the black hole) of the warm emitter derived from the 1st observation were kept frozen for the 2nd observation, which has a shorter exposure (Table~\ref{tbl:obs_log}) and an overall low flux level (Fig.2 of Paper I). That is to say, the warm emitter is assumed to be identical in the two late 2019 observations.  

We note that the photoionizing continuum of the warm emitter is likely different from that of the obscurer and warm absorber. The obscurer is directly exposed to the broadband SED of NGC\,3227. For the warm absorber components, the one with the highest ionization parameter is closest to the central engine (Paper II), thus it is exposed to the filtered broadband SED of NGC\,3227 with the obscurer as the filter. This warm absorber component in turn further filters the photoionizing continuum received by outer and lowly ionized warm absorber components. Hence, all the warm absorber components are de-ionized by the obscurer. 

For the warm emitter, we used the intrinsic continuum derived from the observation taken on 2016-12-05 as its photoionizing continuum. Because distant and/or low-density photoionized plasmas observed as soft X-ray narrow emission lines are likely in a quasi-steady state with their ionization balance \citep{nic99,kaa12,sil16}. 

In Paper I, the intrinsic continuum derived from the observation taken on 2016-12-05 was also used for the reflection component. Here, we performed fits to test alternative photoionization continua. In the first column of Table~\ref{tbl:fit_191115} (Model M1), for the observed spectrum in 2019-11-15, we couple the ionizing continuum of the reflection component to that of the present power law. In the second column (Model M2), we fix the ionizing continuum of the reflection component to that of the power law observed on 2016-12-05, as in Paper I. While the $C$-statistics of two models differ by $\sim28$, their best-fit parameters are consistent with each other within the $3\sigma$ confidence level. For the observed spectrum in 2019-12-05, while most of the best-fit parameters are comparable between Models M1 and M2, the scaling factor\footnote{This scaling parameter (``scale" in Tables~\ref{tbl:fit_191115} and \ref{tbl:fit_191205}) is identical to that of the pexrav model in XSPEC. } of the reflection component does not agree with each other within the $3\sigma$ confidence level (Table~\ref{tbl:fit_191205}). Furthermore, we tested a third model (Model M3) for 2019-12-05, where the ionizing continuum of the reflection component is fixed to that derived from the 2019-11-15 spectrum. In this case, the best-fit continuum parameters are more consistent with those of Model M2 with their $C$-statistics differ by $\sim28$. In these exercises, the choice of the photoionization continuum of the reflection component has no significant impact on the best-fit parameters of the obscurer, which is the main focus of the this work.

\begin{table*}[!h]
\centering
\small
\caption{Best-fit parameters of NGC\,3227 observed on 2019-11-15. The $C$-statistics refer to the final best-fit, where all obscuration, absorption, emission and extinction effects are taken into account. Expected $C$-statistics are calculated as described in \citet{kaa17}.  All quoted errors refer to the statistical uncertainties at the 68.3\% confidence level. Frozen parameters are indicated with (f), which are mainly frozen to values given in Paper I. Coupled parameters are indicated with (c). Both Models M1 and M2 use one photoionized component for the obscurer, while Model M3 uses two photoionized components. The continuum of the reflection component is coupled to that of the power-law component observed on 2019-11-15 for both Models M1 and M3 and 2016-12-05 for Model M2. }
\label{tbl:fit_191115}
\begin{tabular}{l|lll}
\noalign{\smallskip}
\hline
\noalign{\smallskip}
Model & M1 & M2 & M3   \\
\noalign{\smallskip}
\hline
\noalign{\smallskip}
\multicolumn{4}{c}{Statistics} \\
\noalign{\smallskip}
\hline
\noalign{\smallskip}
$C_{\rm stat}$ & 3492.9 & 3465.0 & 3267.5 \\
$C_{\rm expt}$ & $3182\pm81$ & $3182\pm81$ & $3181\pm81$ \\
d.o.f. & 3096 & 3096 & 3093 \\ 
\noalign{\smallskip}
\hline
\noalign{\smallskip}
\multicolumn{4}{c}{Disk blackbody} \\
\noalign{\smallskip}
\hline
\noalign{\smallskip}
Norm (${\rm cm^{-2}}$) & $4.9\times10^{26}$ (f) & $4.9\times10^{26}$ (f) & $4.9\times10^{26}$ (f)  \\
$T$ (eV) & $10.2$ (f) & $10.2$ (f) & $10.2$ (f)\\
\noalign{\smallskip}
\hline
\noalign{\smallskip}
\multicolumn{4}{c}{Comptonisation} \\
\noalign{\smallskip}
\hline
\noalign{\smallskip}
Norm (${\rm ph~s^{-1}~keV^{-1}}$) & $(2.0\pm0.1)\times10^{53}$ & $(2.1\pm0.1)\times10^{53}$ & $(4.9\pm0.8)\times10^{53}$  \\
$T_{\rm seed}$ (eV) & $10.2$ (c) & $10.2$ (c) & $10.2$ (c)\\
$T_{\rm c}$ (keV) & $0.06$ (f) & $0.06$ (f) & $0.06$ (f) \\
$\tau$ & $30$ (f) & $30$ (f) & $30$ (f) \\
\noalign{\smallskip}
\hline
\noalign{\smallskip}
\multicolumn{4}{c}{Power-law} \\
\noalign{\smallskip}
\hline
\noalign{\smallskip}
Norm (${\rm ph~s^{-1}~keV^{-1}}$) & $(2.99\pm0.06)\times10^{50}$  & $(2.85\pm0.06)\times10^{50}$ & $(3.73\pm0.10)\times10^{50}$   \\
$\Gamma$ & $1.74\pm0.01$ & $1.72\pm0.01$ & $1.83\pm0.01$ \\
\noalign{\smallskip}
\hline
\noalign{\smallskip}
\multicolumn{4}{c}{Reflection} \\
\noalign{\smallskip}
\hline
\noalign{\smallskip}
Norm (${\rm ph~s^{-1}~keV^{-1}}$) & $2.99\times10^{50}$ (c) & $3.87\times10^{50}$ (f) & $3.73\times10^{50}$ (c) \\
$\Gamma$ & $1.74$ (c) & $1.83$ (f) & 1.83 (c) \\
scale & $0.44\pm0.03$ & $0.42\pm0.02$ & $0.47\pm0.03$ \\
\noalign{\smallskip}
\hline
\noalign{\smallskip}
\multicolumn{4}{c}{Obscurer}  \\
\noalign{\smallskip} 
\hline
\noalign{\smallskip}
$N_{\rm H}~({\rm 10^{22}~cm^{-2}})$ & $4.0\pm0.1$ & $3.8\pm0.1$ & $9.2\pm0.6$, $1.0\pm0.2$ \\
\noalign{\smallskip}
$\log \xi~{\rm (erg~s^{-1}~cm)}$ & $0.5\pm0.3$ & $0.5\pm0.2$ & $<2.0$, $-0.7_{-0.4}^{+0.3}$ \\
\noalign{\smallskip}
$f_{\rm cov}^{\rm X}$ & $0.610\pm0.008$ & $0.595\pm0.008$ & $0.44\pm0.02$, $0.60\pm0.02$ \\
\noalign{\smallskip}
\hline
\noalign{\smallskip}
\multicolumn{4}{c}{Warm emitter} \\
\noalign{\smallskip}
\hline
\noalign{\smallskip}
$N_{\rm H}~({\rm 10^{21}~cm^{-2}})$ & $9.6\pm1.6$ & $9.8\pm1.6$ & $11.7\pm1.7$ \\
\noalign{\smallskip}
$\log \xi~{\rm (erg~s^{-1}~cm)}$ & $1.37\pm0.12$ & $1.38\pm0.12$ & $1.51\pm0.09$ \\
\noalign{\smallskip}
$C_{\rm em}~(\Omega/4\pi)$ & 0.01 (f) & 0.01 (f) & 0.01 (f) \\
\noalign{\smallskip}
$v_{\rm mac}~{\rm (km~s^{-1})}$ & $950_{-240}^{+320}$ & $960_{-240}^{+300}$ & $1070_{-260}^{+330}$ \\ 
\noalign{\smallskip}
\hline
\end{tabular}
\end{table*}

\begin{table*}[!h]
\centering
\small
\caption{Similar to Table~\ref{tbl:fit_191115} but for 2019-12-05. Models M1, M2 and M3 use one photoionized component for the obscurer, while Model M4 uses two photoionized components. The continuum of the reflection component is coupled to that of the power-law component observed on 2019-12-05 for Model M1 and M4, 2016-12-05 for Model M2 (as in Paper I), and 2019-11-15 for Model M3. }
\label{tbl:fit_191205}
\begin{tabular}{l|llll}
\noalign{\smallskip}
\hline
\noalign{\smallskip}
Model & M1 & M2 & M3 & M4  \\
\noalign{\smallskip}
\hline
\noalign{\smallskip}
\multicolumn{5}{c}{Statistics} \\
\noalign{\smallskip}
\hline
\noalign{\smallskip}
$C_{\rm stat}$ & 3600.1 & 3580.5 & 3608.3 & 3497.5 \\
$C_{\rm expt}$ & $3383\pm81$ & $3384\pm81$ & $3384\pm81$ & $3388\pm81$ \\
d.o.f. & 3054 & 3054 & 3054 & 3051 \\ 
\noalign{\smallskip}
\hline
\noalign{\smallskip}
\multicolumn{5}{c}{Disk blackbody} \\
\noalign{\smallskip}
\hline
\noalign{\smallskip}
Norm (${\rm cm^{-2}}$) & $4.0\times10^{26}$ (f) & $4.0\times10^{26}$ (f) & $4.0\times10^{26}$ (f) & $4.0\times10^{26}$ (f)  \\
$T$ (eV) & $10.2$ (f) & $10.2$ (f) & $10.2$ (f) & $10.2$ (f)\\
\noalign{\smallskip}
\hline
\noalign{\smallskip}
\multicolumn{5}{c}{Comptonisation} \\
\noalign{\smallskip}
\hline
\noalign{\smallskip}
Norm (${\rm ph~s^{-1}~keV^{-1}}$) & $1.1_{-1.1}^{+1.3}\times10^{52}$ & $4.8_{-2.4}^{+1.3}\times10^{52}$ & $3.1_{-1.2}^{+2.5}\times10^{52}$ & $18.3_{-5.6}^{+7.1}\times10^{52}$ \\
$T_{\rm seed}$ (eV) & $10.2$ (c) & $10.2$ (c) & $10.2$ (c) & $10.2$ (c)\\
$T_{\rm c}$ (keV) & $0.06$ (f) & $0.06$ (f) & $0.06$ (f) & $0.06$ (f)\\
$\tau$ & $30$ (f) & $30$ (f) & $30$ (f) & $30$ (f) \\
\noalign{\smallskip}
\hline
\noalign{\smallskip}
\multicolumn{5}{c}{Power-law} \\
\noalign{\smallskip}
\hline
\noalign{\smallskip}
Norm (${\rm ph~s^{-1}~keV^{-1}}$) & $(9.0\pm0.4)\times10^{49}$  & $(6.7\pm0.4)\times10^{49}$ & $(7.3\pm0.5)\times10^{49}$ & $(22.4\pm3.7)\times10^{49}$ \\
$\Gamma$ & $1.82\pm0.03$ & $1.69\pm0.03$ & $1.70\pm0.03$ & $2.02\pm0.04$ \\
\noalign{\smallskip}
\hline
\noalign{\smallskip}
\multicolumn{5}{c}{Reflection} \\
\noalign{\smallskip}
\hline
\noalign{\smallskip}
Norm (${\rm ph~s^{-1}~keV^{-1}}$) & $9.0\times10^{49}$ (c) & $38.7\times10^{49}$ (f) & $29.9\times10^{49}$ (f) & $22.4\times10^{49}$ (c) \\
$\Gamma$ & $1.82$ (c) & $1.83$ (f) & 1.74 (f) & 2.02 (c) \\
scale & $1.41\pm0.14$ & $0.30\pm0.02$ & $0.29\pm0.02$ & $0.83\pm0.11$ \\
\noalign{\smallskip}
\hline
\noalign{\smallskip}
\multicolumn{5}{c}{Obscurer}  \\
\noalign{\smallskip} 
\hline
\noalign{\smallskip}
$N_{\rm H}~({\rm 10^{22}~cm^{-2}})$ & $10.5\pm0.1$ & $7.0_{-0.8}^{+2.7}$ & $9.4_{-2.6}^{+1.2}$ & $56_{-6}^{+8}$, $4.1_{-0.6}^{+1.3}$ \\
\noalign{\smallskip}
$\log \xi~{\rm (erg~s^{-1}~cm)}$ & $2.6\pm0.2$ & $2.0\pm0.6$ & $2.5_{-0.8}^{+0.1}$ & $0.5_{-1.0}^{+1.2}$, $-0.5_{-3.8}^{+2.4}$ \\
\noalign{\smallskip}
$f_{\rm cov}^{\rm X}$ & $0.72\pm0.02$ & $0.64\pm0.04$ & $0.67\pm0.04$ & $0.54\pm0.05$, $0.75\pm0.02$ \\
\noalign{\smallskip}
\hline
\end{tabular}
\end{table*}

Between the two observations in 2019 fitted with Model M1, we found that while the $1-1000$~Ryd ionizing flux was lower in December (smaller by a factor of $\sim2$), the ionization parameters ($\xi$) were larger (larger by a factor of $\gtrsim100$). If $n_{\rm H}r^2$ of the obscurer remains the same in December, one would expect the ionization parameter to be smaller by a factor of $\sim2$ according to the definition of the ionization parameter. The best-fit ionization parameter does not support a constant $n_{\rm H}r^2$. That is to say, unless the number density and/or distance of the obscurer decreases significantly within one month, the observed data cannot be explain by the same obscurer. The hydrogen column density is also larger by a factor of $\sim2$ in December. Therefore, we might catch two different obscurers in the two observations in 2019. A detailed variability study of the obscurer as well as the intrinsic continuum is beyond the scope of this work. We refer readers to Paper IV (Grafton-Waters et al. in prep.). 

For both 2019 spectra, we also attempted to add another \textit{pion} component for the obscurer. The best-fit results are listed as Model M3 in Table~\ref{tbl:fit_191115} and Model M4 in Table~\ref{tbl:fit_191205}. When using two \textit{pion} components instead of one, the $C$-statistics can be significantly improved. Between the two obscuring components, the hydrogen column densities ($N_{\rm H}$) differed by nearly an order of magnitude. The one with relatively large $N_{\rm H}$ is the additional component.
For 2019-11-15, the ionization parameter of the leading $N_{\rm H}$ component cannot be well constrained. For 2019-12-05, the ionization parameters of both components were poorly constrained. Moreover, the hydrogen column density of the additional photoionized component ($\log \xi \sim0.5$) is much higher than the other component. The significantly increase normalization of the continuum components also contributes to the increased $N_{\rm H}$ for the additional component. Without observable discrete absorption lines of the obscurer, adding a second \textit{pion} component introduces degeneracy among parameters ($N_{\rm H}$, $\log \xi$, and the normalizations of the continuum components). Thus, these parameters can have relatively large uncertainties. 

\begin{figure*}
\centering
\includegraphics[width=.8\hsize, trim={0.cm 0.5cm 2.cm 0.0cm}, clip]{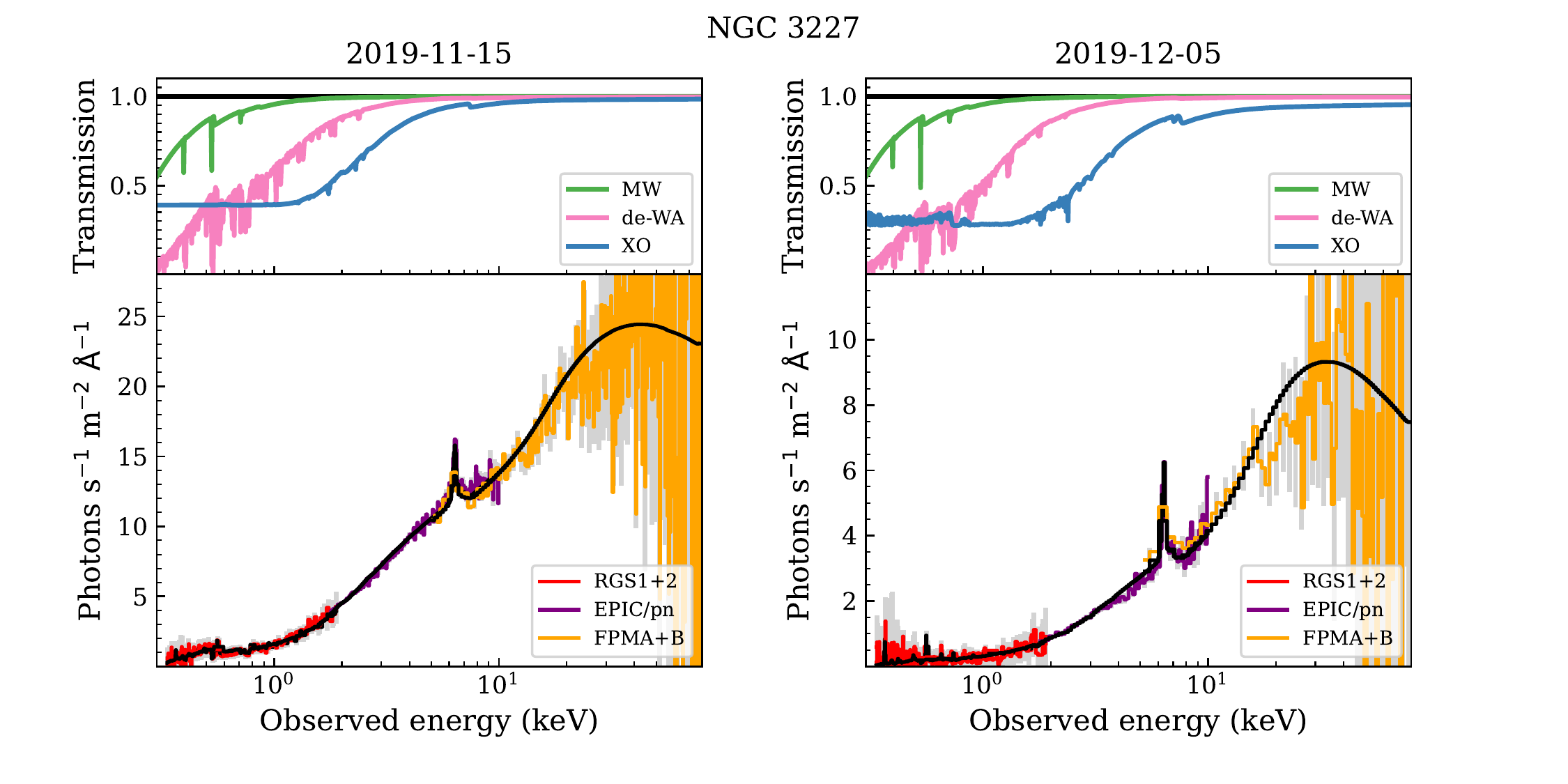}
\caption{The best-fit model (Model M1 in Tables~\ref{tbl:fit_191115} and \ref{tbl:fit_191205}) to the XMM-\textit{Newton} (EPIC/pn and RGS) and NuSTAR spectra of NGC\,3227 in late 2019. The top panels are the transmission of the X-ray obscurer (XO), de-ionized warm absorber (de-WA), and the Galactic absorption (MW). Data (colored curves with $1\sigma$ uncertainties in gray) and model (black curves) of each instrument are rebinned for clarity.  } 
\label{fig:plot_dmtk_2019}
\end{figure*}

\begin{figure*}
\centering
\includegraphics[width=.8\hsize, trim={0.cm 0.cm 2.cm 0.0cm}, clip]{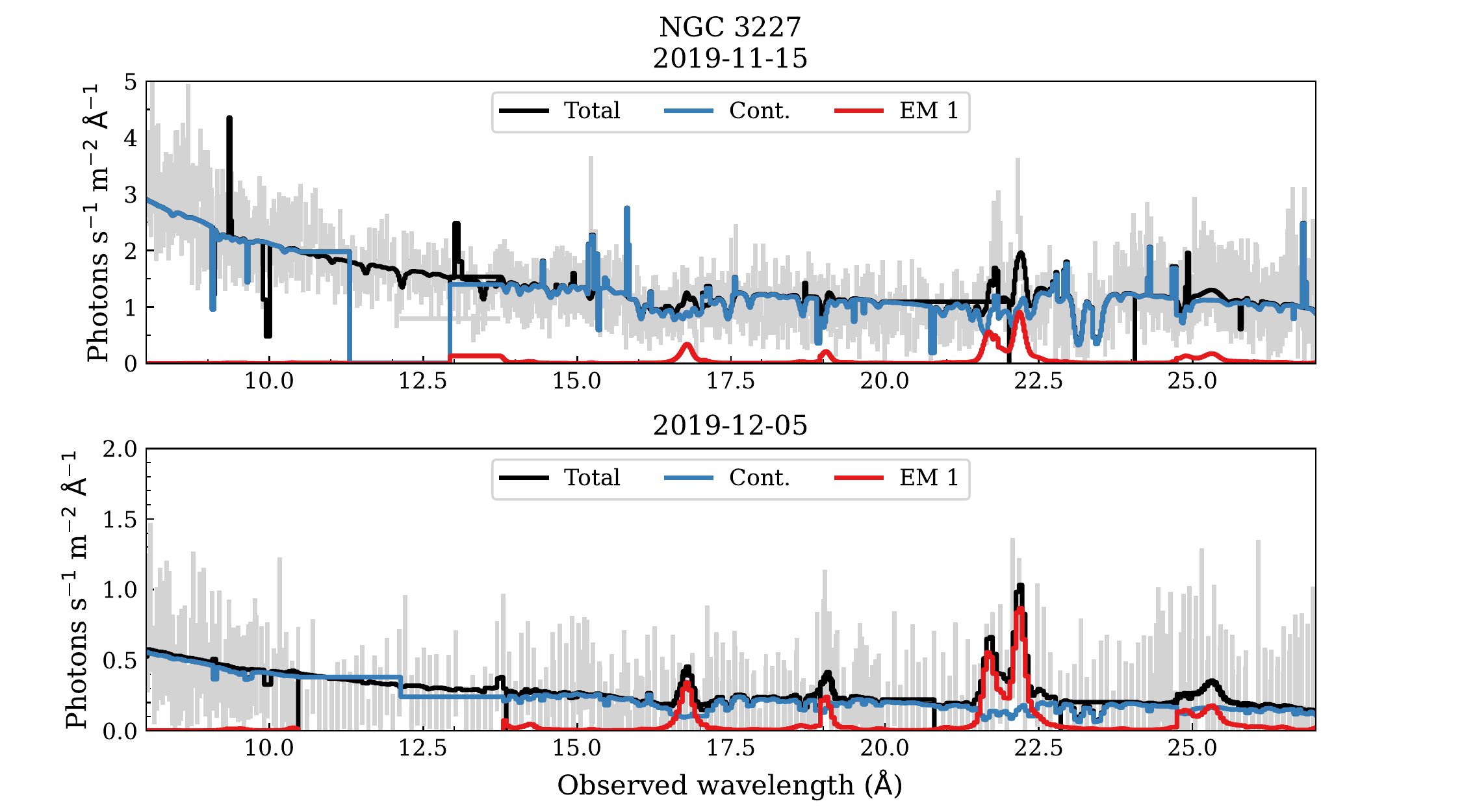}
\caption{The best-fit model (Model M1 in Tables~\ref{tbl:fit_191115} and \ref{tbl:fit_191205}) to the RGS spectra of NGC\,3227 on 2019-11-15 (top) and 2019-12-05 (bottom). The warm emitter component is shown in red and identical for both observations, while the continuum is shown in blue. The black curves are the total emission. } 
\label{fig:plot_cf_dnma}
\end{figure*}

As discussed later, we are not able to pinpoint the location of the X-ray obscurer. We are also puzzled by the lack of UV absorption features of the X-ray obscurer. Using two photoionized components instead of one for the X-ray obscurer does not mitigate these issues. For simplicity, we only show the best-fit model to the observed data for Model M1 in Fig.~\ref{fig:plot_dmtk_2019}. In the same figure, the transmissions of the obscurer, de-ionized warm absorber, and Galactic absorption can also be found. Fig.~\ref{fig:plot_cf_dnma} shows the best-fit model to the observed RGS data in the soft X-ray band. Note that the flux level of the 2019-12-05 spectrum is rather low and the emission lines of the warm emitter are barely observed. 

\begin{figure*}
\centering
\includegraphics[width=\hsize, trim={0.5cm 0.5cm 0.5cm 0.0cm}, clip]{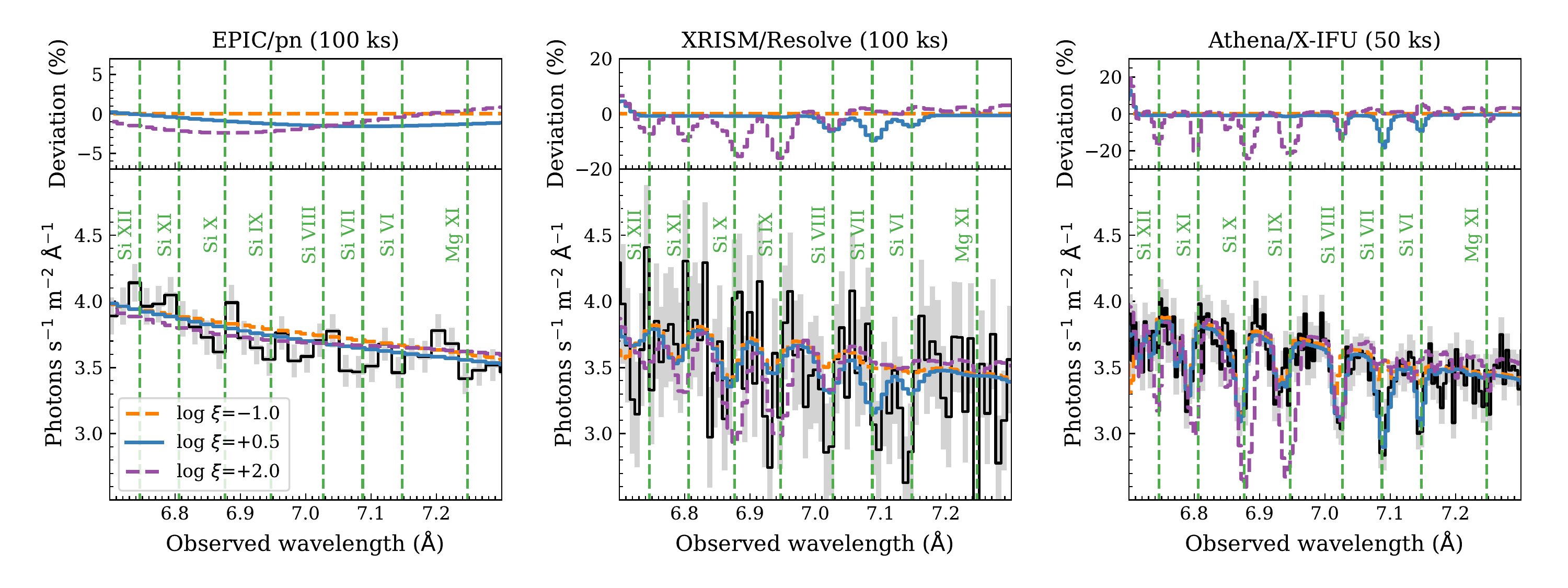}
\caption{X-ray spectra around $\sim7.0$~\AA\ (or $\sim1.77$~keV) for EPIC/pn (100 ks, left),  XRISM/Resolve (100 ks, middle), and Athena/X-IFU (50 ks, right). Data are shown in black with $1\sigma$ uncertainties shown in gray. Absorption lines from Si {\sc vi} to Si {\sc xii} and Mg {\sc xi} are highlighted with vertical dashed lines in green. Three sets of models with different ionization parameters for the X-ray obscurer are shown: $\log \xi=-1.0$ in orange and dashed lines, $\log \xi=0.5$ in blue and solid lines, $\log \xi=2.0$ in purple and dashed lines, respectively. Deviation (in percentage) among the three models are shown in the upper panels.} 
\label{fig:plot_cf_dnma_inst}
\end{figure*}

We caution that the transmission plots (upper panels of Fig.\ref{fig:plot_dmtk_2019}) are derived from plasma models in a rather fine energy grid. Si and S absorption lines around $\sim1.5-3$~keV are not resolved with the current instruments. In the left panel of Fig.\ref{fig:plot_cf_dnma_inst}, we show the spectral region around $\sim7.0$~\AA\ (or $\sim1.77$~keV) for EPIC/pn. Si {\sc vi} to Si {\sc xii} absorption lines from the obscurer dominates this energy range, probing a wide range in ionization parameter \citep[e.g.,][]{mao17}. Three sets of models with different ionization parameters for the X-ray obscurer are shown for comparison. Based on Model M1 in Table~\ref{tbl:fit_191115}, we re-fitted the observed 2019-11-15 data set with different ionization parameters (frozen). Although Si lines are not resolved with EPIC/pn, models with different ionization parameters can lead to different C-statistics, e.g., $\Delta C\sim9$ between $\log \xi=-1.0$ and $\log \xi=0.5$, in this narrow wavelength range of $6.7-7.3$~\AA. This explains the relatively small $1\sigma$ ($\Delta C=1$) statistical uncertainties for the obscurer in Tables~\ref{tbl:fit_191115} and \ref{tbl:fit_191205}. In Fig.\ref{fig:plot_cf_dnma_inst}, we also show simulated XRISM/Resolve \citep{xri20} and Athena/X-IFU \citep{bar18} spectra. The Si absorption lines can be better resolved with XRISM/Resolve, which has a relatively small effective area though. Athena/X-IFU is the most ideal instrument to put tight constraints on the ionization parameter of the X-ray obscurer.

\begin{figure}
\centering
\includegraphics[width=.9\hsize, trim={0.cm 0.cm .cm 0.0cm}, clip]{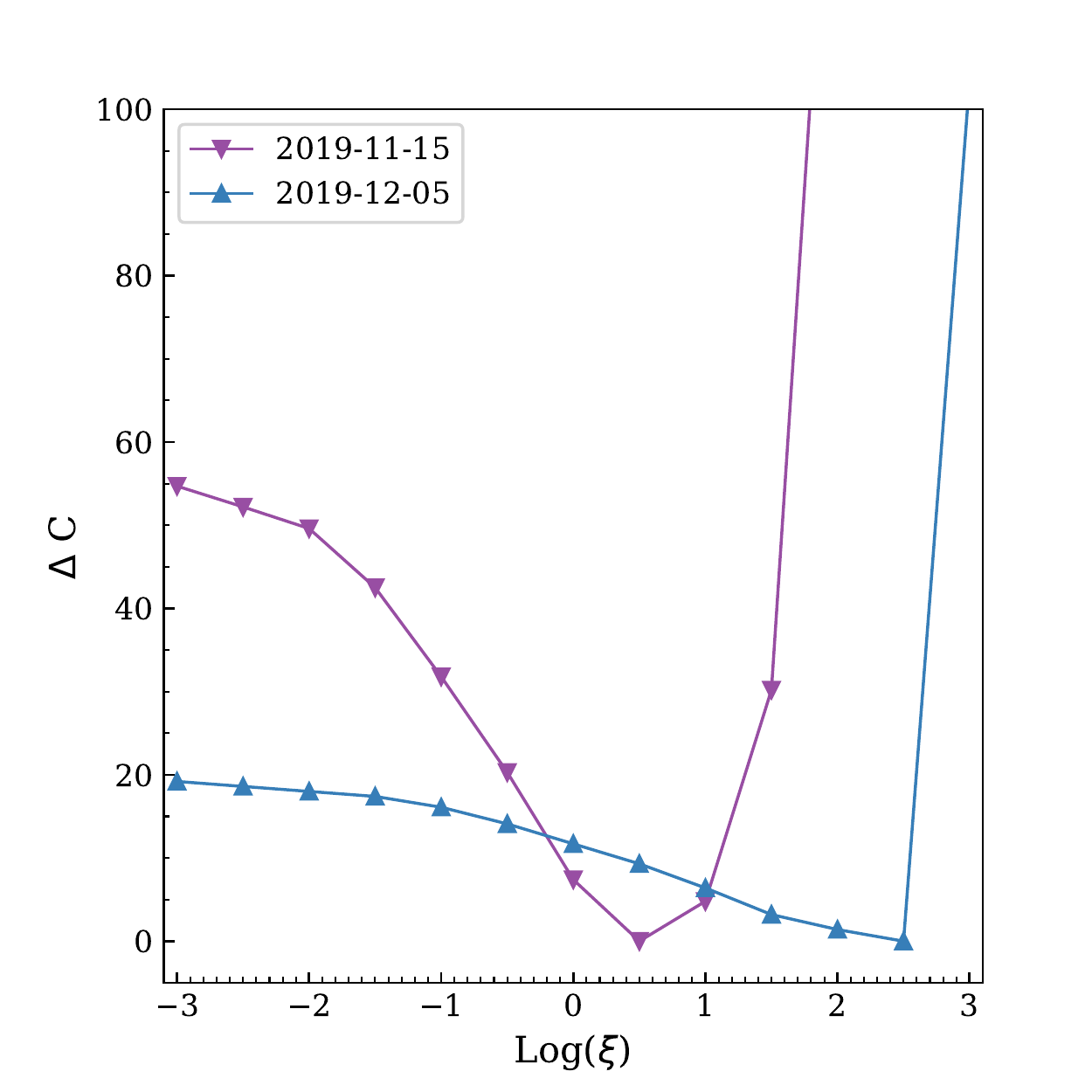}
\caption{The change of $C$-statistics ($\Delta C$) for models with different ionization parameters for the X-ray obscurer. These are obtained by varying the ionization parameter of the obscurer (frozen to different values) of Models M1 in Table~\ref{tbl:fit_191115} and \ref{tbl:fit_191205} and re-fit the observed 2019 spectra.} 
\label{fig:plot_cf_cstat}
\end{figure}

With the above being said, we emphasize that the current data sets can rule out a highly ionized X-ray obscurer. Based on Model M1 for both observations, we fixed the ionization parameter of the obscurer to different values spanning six decades and re-fit the observed spectra. The changes of $C$-statistics ($\Delta C$) are shown in Fig.~\ref{fig:plot_cf_cstat}, where $\Delta C$ is rapidly increasing when $\log \xi \gtrsim1.5$ for 2019-11-15 and $\log \xi \gtrsim3.0$ for 2019-12-05. On one hand, this is due to the lack of absorption lines (e.g., Si {\sc x}, S {\sc xv}, Fe {\sc xxvi}) in the observed spectra. On the other hand, a large fraction of soft X-ray photons would leak through a highly ionized obscurer. Taking into account the absorption effect of the de-ionized warm absorber, this would lead to a rather different continuum shape in the soft X-ray band than the observed one. A lowly ionized X-ray obscurer is also statistically unacceptable, especially for the 2019-11-15 data set. Due to the relatively short exposure (Table~\ref{tbl:obs_log}) and low flux (e.g., Fig.~\ref{fig:plot_dmtk_2019}), the 2019-12-05 data set is less sensitive ($\Delta C\lesssim20$) to a wide range of ionization parameters. 

\section{C {\sc iv} emission line profiles}
\label{sct:030604_lpro}
For the long-lasting X-ray obscuration event in NGC 5548 \citep{kaa14}, fingerprints of the obscurer were found as blueshifted broad absorption troughs in the UV \citep{kri19b} and NIR \citep{lan19,wil21} collected in $2011-2016$. Since the obscurer is closer to the central engine than the warm absorber, the shielding effect provided by the obscurer can also give rise to new narrow absorption lines of lowly-ionized species of the warm absorber, as well as variation of the existing narrow UV absorption lines of the warm absorber \citep{ara15}. Blueshifted broad absorption troughs in the UV spectra were also found for the short-lived X-ray obscuration events in December 2016 for NGC\,3783 \citep{meh17}. Note that X-ray obscuration events in NGC\,3783 are also recurrent \citep{kaa18}.

\begin{figure*} 
\centering
\includegraphics[width=.9\hsize, trim={1.cm 0.cm .cm .0cm}, clip]{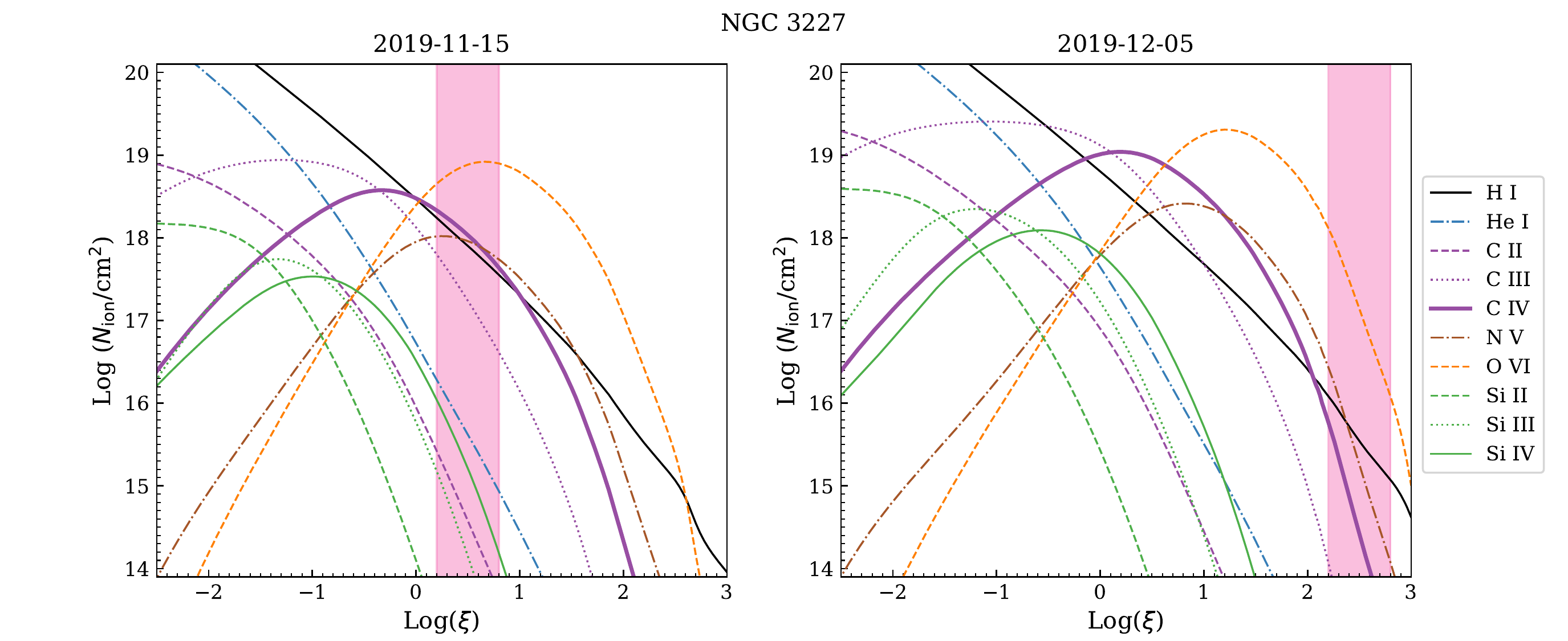}
\caption{Ionic column densities of the obscurer in NGC\,3227 observed on 2019-11-15 (left) and 2019-12-05 (right). The shaded areas mark the $1\sigma$ range of the ionization parameters of the obscurer (Model M1 in Tables~\ref{tbl:fit_191115} and \ref{tbl:fit_191205}).} 
\label{fig:plot_cold_ow}
\end{figure*}

For NGC\,3227, if the obscurer observed in the X-ray band intercepts our line of sight to (a significant portion of) the UV emitting region, we would expect to see absorption features in the UV band. As shown in Fig.~\ref{fig:plot_cold_ow}, the ionic column densities of e.g., H {\sc i}, C {\sc iv}, N {\sc v}, and Si {\sc iv} for the X-ray obscurer are all well above $10^{14}~{\rm cm^{-2}}$ for a wide range of the ionization parameter ($-3\lesssim \log \xi \lesssim 3$). These column densities are large enough to produce blueshifted absorption troughs in the HST/COS spectra for NGC\,3227, similar to those in NGC 5548 \citep{kaa14} and NGC 3783 \citep{meh17, kri19a}. Although a highly ionized X-ray obscurer can result in low ionic column densities, it is not consistent with the observed X-ray data (Sect.~\ref{sct:spec}).

\begin{figure}
\centering
\includegraphics[width=.9\hsize, trim={0.cm 0.cm 1.cm .0cm}, clip]{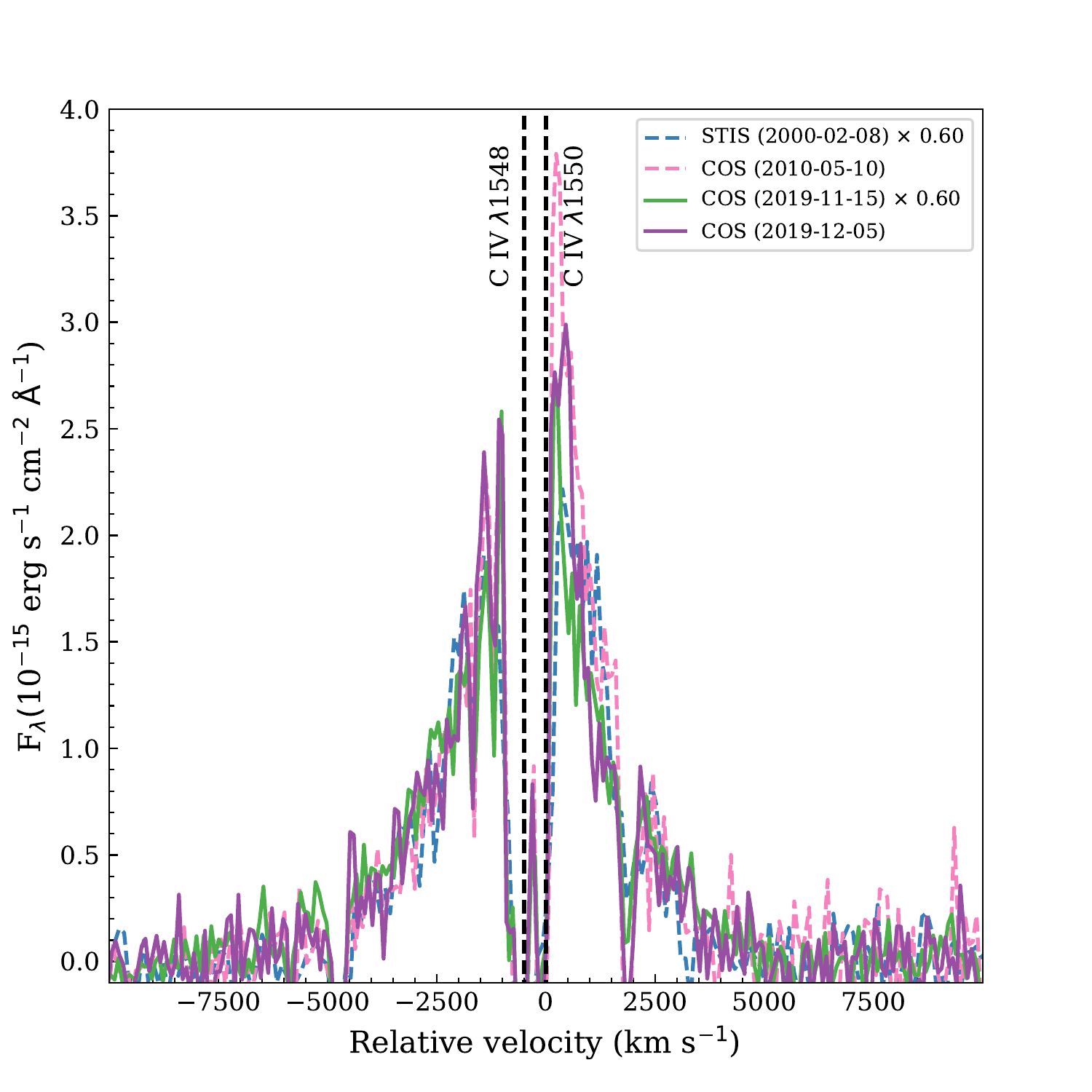}
\caption{C {\sc iv} line profile in NGC\,3227 observed with HST/COS on 2000-02-08 (blue), 2010-05-10 (pink), 2019-11-15 (green), and 2019-12-05 (purple). Local continua are subtracted here. Both the 2000-02-08 and 2019-11-15 line profiles are scaled by 0.6. The vertical dashed lines mark the C {\sc iv} $\lambda\lambda1548,1550$ doublet.
} 
\label{fig:plot_cf_velspec_uv_040603}
\end{figure}

Fig.~\ref{fig:plot_cf_velspec_uv_040603} compares the two 2019 HST/COS spectra with the 2010-05-10 HST/COS and 2000-02-08 HST/STIS spectra. No prominent blueshifted broad absorption troughs were found in any of these data. We did not find the emergence of new narrow absorption features either. No significant variations among the known narrow absorption lines were found although we caution that the strongest ones were saturated. The apparent variable feature at $\sim+1800~{\rm km~s^{-1}}$ in the C {\sc iv} emission-line profile is Si {\sc i} $\lambda1568$ absorption intrinsic to NGC\,3227. Its actual variability is not significant. It appears enhanced in Fig.~\ref{fig:plot_cf_velspec_uv_040603} due to the continuum subtraction and scaling applied to individual spectra so that the blue wing of C {\sc iv} are comparable. 

Multiple X-ray obscuration events have been found in NGC\,3227: $2000-2001$ \citep{lam03,mar14}, 2002 \citep{mar14}, 2006 (Paper II), 2008 \citep{beu15}, 2016 \citep[][and Paper II]{tur18}, and 2019 (present work). We first check whether NGC\,3227 was obscured in the X-ray band on 2000-02-08 and 2010-05-10. If so, it might explain the similarity of the scaled C {\sc iv} emission line profiles. 

Unfortunately, no X-ray observations were available in the entire year of 2010. On the other hand, there was a weekly monitoring campaign of NGC\,3227 with RXTE in early 2000 \citep{mar14}. According to the authors, between November 2000 and February 2001, the general indicator of hard X-ray spectral shape (``apparent photon index" introduced by the authors) varied significantly, similar to the two secure obscuration events identified (their Fig. A12). Moreover, the maximum hardness ratio of $F_{\rm 7-10~keV}/F_{\rm 2-4~keV}$ was $6\sigma$ above the average. Nonetheless, this period was not considered as an obscuration event because it failed to meet one of the criteria of secure events defined by \citet{mar14}, where the hardness ratio was required to increase for at least two consecutive snapshots in a row. 

We cannot rule out the possibility that a short-lived (less than the 1-week cadence of RXTE observations in early 2000) obscuration event occurred on 2000-02-08. \citet{mar14} identified a secure short ($\sim2.1-6.6$~day) obscuration event with RXTE in Oct. 2002. On 2016-12-09, another short-lived ($\lesssim4$~d) obscuration event was identified \citep[][and Paper II]{tur18} with deep XMM-Newton observations. Note that XMM-Newton provides data in the soft X-ray band below 2 keV, which is outside the band pass of RXTE but the soft X-ray band is where the obscuration effect is most prominent (e.g., Figure 2 of Paper I). 

Considering the variation (on both shorter and longer timescales) of the known obscurer in other Seyfert galaxies \citep{dge15,meh16a,cap16,dma20}, even if X-ray obscuration events occurred on 2000-02-08 and 2010-05-10, it is still difficult to reconcile the similar UV absorption features in all four HST spectra. Therefore, we explore possible explanations of the lack of new UV absorption features assuming that X-ray obscuration events were absent on 2000-02-08 and 2010-05-10.

\section{Discussion}
\label{sct:dis}
We argue that large ionic column densities inferred from the X-ray analysis (Fig.~\ref{fig:plot_cold_ow}) do not necessarily produce observable absorption features in HST/COS. This might be explained if the X-ray obscurer does not intercept our line of sight to (a significant portion of) the UV emitting region. There are two possible scenarios: (1) the obscurer is launched in the vicinity of the central engine and has not reached the UV emitting region yet; (2) the obscurer is above the UV emitting region but it is rather compact in size so that it does not cover a significant portion of the UV emitting region. For the former, we estimate the distance of the obscurer and compared it to the UV emitting region (Sect.~\ref{sct:dist2BH}). For the latter, we showed this is plausible considering the size of the X-ray and UV emitting region and the X-ray covering factor (Sect.~\ref{sct:uv_fcov}).

\subsection{Effective UV emitting region radius}
\label{sct:dist_uv}
We first estimate the effective UV emitting region radius ($R_{2500~\AA}$) following \citet[][Eq. S7]{bur21}
\begin{equation}
    R_{2500~\angstrom} = 10^{14.95\pm0.05}~{\rm cm}~\left(\frac{L_{5100~\angstrom}}{10^{44}~{\rm erg~s^{-1}}}\right)^{0.53\pm0.04}~, 
\end{equation}
where $L_{5100~\AA}$ is the optical continuum luminosity. For NGC\,3227, with $L_{5100~\AA}\sim3\times10^{42}~{\rm erg~s^{-1}}$ \citep{dro15}, we have $R_{\rm 2500~\AA}\sim1.4\times10^{14}~{\rm cm}$ or $\sim0.05$~ld. This is equivalent to $\sim80~R_{\rm S}$, where the Schwarzschild radius $R_{\rm S}=2GM_{\rm BH}/c^2=1.76\times10^{12}~{\rm cm}$. 

\subsection{Distance estimation of the X-ray obscurer}
\label{sct:dist2BH}
We estimate the distance of the obscurer to the central engine and compare it to the distance of the BLR and torus given in the literature. The distance estimation is based on the following assumptions. The obscurer has a uniform density and ionization parameter $\xi=L_{\rm ion}/n_{\rm H}~r^2$ \citep{tar69,kro81}, where $L_{\rm ion}$ is the $1-1000$ Ryd ionizing luminosity, $n_{\rm H}$ is the hydrogen number density of the obscurer, $r$ is the distance to the central engine. The length scale of the obscurer along the line of sight is $N_{\rm H}/n_{\rm H}$. We assume that the length scale of the obscurer across the line of sight is simply $fN_{\rm H}/n_{\rm H}$, where $f$ is the ratio of the azimuthal to radial length scale. The obscurer might be viewed as a stream line ($f<1$), a spherical cloud ($f=1$), or a flat bread ($f>1$). The velocity of the obscurer across our line of sight is $v_{\rm cross}=f~N_{\rm H}/(n_{\rm H}~t_{\rm cross})$, where $t_{\rm cross}$ is the crossing time. We further assume $v_{\rm cross}=\sqrt{G~M_{\rm BH}/r}$, where $G$ is the gravitational constant and $M_{\rm BH}$ the black hole mass. Note that the radial velocity can be much larger than the crossing (or azimuthal) velocity. We have 
\begin{equation}
\label{eq:dist2BH}
    r\simeq \left(\frac{15.4}{\rm ld}\right)~M_{\rm BH,7}^{1/5} \left(\frac{L_{\rm ion,42}}{f~N_{\rm H,22}~\xi}~\frac{t_{\rm cross}}{\rm day}\right)^{2/5}~,
\end{equation}
where $M_{\rm BH,7}=M_{\rm BH}/10^7~M_{\odot}$, $L_{\rm ion, 42}=L_{\rm ion}/10^{42}~{\rm erg~s^{-1}}$, and $N_{\rm H,22}=N_{\rm H}/10^{22}~{\rm cm^{-2}}$. With $f=1$, we obtain the distance estimation equations used by \citet{lam03} and \citet{beu15}. 

Table~\ref{tbl:dist2BH} lists the distance estimation of the obscurer to the black hole for $f=0.1$, 1, 10, and 3000, including all the previously identified X-ray obscuration events \citep[][and Paper II]{lam03,mar14,beu15,tur18}. Originally, \citet[][M14]{mar14} used a black hole mass of $7.59\times10^6~M_{\odot}$ \citep{den10} and \citet[][B15]{beu15} used the average value of various black hole mass measurements of $1.75\times10^7~M_{\odot}$. Following Paper I, we adopt the black hole mass of $5.96\times10^6~M_{\odot}$ from \citet{ben15}. The distance to the black hole for the X-ray obscuration events identified by \citet{mar14} and \citet{beu15} are re-calculated for comparison purposes. 

\begin{sidewaystable*}[h]
\centering
\caption{Distance of the obscurers to the black hole (Eq.~\ref{eq:dist2BH}) with $M_{\rm BH}=5.96\times10^6~M_{\odot}$ \citep{ben15}. M14 and B15 are short for \citet{mar14} and \citet{beu15}, respectively. Distances are calculated for three representative geometries: a stream line ($f=0.1$), a spherical cloud ($f=1$), and a flat bread ($f=10$). When multiple obscuration events occur on the same day, we use lower case letters (a and b here) to differentiate them. Caution that the ionization parameter of 2019-12-05 (L), marked with $\dagger$, is poorly constrained (Table~\ref{tbl:fit_191205}). }
\label{tbl:dist2BH}
\begin{tabular}{lccccccccccccc}
\noalign{\smallskip}
\hline
\noalign{\smallskip}
Event & $2000-2001$ & 2002 & 2006-12-03 & 2008 & 2016-12-09a & 2016-12-09b & 2019-11-15 & 2019-12-05 \\
\noalign{\smallskip}
\hline
\noalign{\smallskip}
Ref. & M14 & M14 & Paper II & B15 & Paper II & Paper II & Paper III & Paper III \\
\noalign{\smallskip}
$L_{\rm ion}$ $({\rm erg~s^{-1}})$ & $10^{43}$ & $10^{43}$ & $19.1\times10^{42}$ & $8.9\times10^{42}$ & $23.5\times10^{42}$ & $23.5\times10^{42}$ & $11.7\times10^{42}$ & $5.8\times10^{42}$ \\
\noalign{\smallskip}
$t_{\rm cross}$ (day) & $77-94$ & $13.3$ & $0.23-372$ & $\gtrsim35$ & $0.45-3.53$ & $\sim0.23$  & $1.2-50$ & $\gtrsim0.58-127$ \\
\noalign{\smallskip}
$N_{\rm H}$ $(10^{22}~{\rm cm^{-2}})$ & $19-26$ & 13.3 & 1.98 & 11.2 & 8.27, 1.25 & 1.33 & 3.5 & 8.3 \\
\noalign{\smallskip}
$\log \xi$ & $0.0,~-0.3$ & $0.0$ & $1.55$ & $1.1$ & $2.81,~1.02$ & 1.89 & 0.4 & 2.0 \\
\noalign{\smallskip}
\hline
\noalign{\smallskip}
\multicolumn{9}{c}{$f=0.1$ (stream line)} \\
\noalign{\smallskip}
\hline
\noalign{\smallskip}
$r$ (ld) & $144-156$, $189-205$ & $42-66$ & $12-222$ & $48-73$ & $2.9-6.7$, $32-73$ & $\sim11$ & $36-162$ & $2.0-17$  \\ 
$n_e~(10^8~{\rm cm^{-3}})$ & $0.6-0.7$, $0.7-0.8$ & $3.4-8.5$ & $0.02-6.0$ & $0.2-0.5$ & $1.2-6.5$, $0.6-3.2$ & $\sim3.9$ & $0.2-4.3$ & $0.07-5.3$ \\
Radial size ($10^2~R_S$) & $18-21$ , $15-18$ & $0.9-2.2$ & $0.2-69$ & $14-32$ & $0.7-3.8$ , $0.2-1.1$ & $\sim0.2$ & $0.5-11$ & $1.1-84$   \\
\noalign{\smallskip}
\hline
\noalign{\smallskip}
\multicolumn{9}{c}{$f=1$ (spherical cloud)} \\
\noalign{\smallskip}
\hline
\noalign{\smallskip}
$r$ (ld) & $57-62$, $75-82$ & $17-26$ & $4.6-88$ & $19-29$ & $1.2-2.7$, $13-29$ & $\sim4.3$ & $14-65$ & $0.8-7.0$ \\
$n_e~(10^9~{\rm cm^{-3}})$ & $0.4-0.5$, $0.4-0.5$ & $2.1-5.3$ & $0.01-3.8$ & $0.1-0.3$ & $0.8-4.1$, $0.4-2.0$ & $\sim2.5$ & $0.1-2.7$ & $0.04-3.4$ \\
Radial size ($R_S$) & $281-329$, $245-287$ & $14-35$ & $3.0-1092$ & $221-511$ & $12-60$, $3.5-18$ & $\sim3.1$ & $8.4-172$ & $18-1328$  \\
\noalign{\smallskip}
\hline
\noalign{\smallskip}
\multicolumn{9}{c}{$f=10$ (flat bread)} \\
\noalign{\smallskip}
\hline
\noalign{\smallskip}
$r$ (ld) & $23-25$, $30-33$ & $6.7-11$ & $1.8-35$ & $7.6-12$ & $0.5-1.1$, $5.1-12$ & $\sim1.7$ & $5.7-26$ & $0.3-2.8$  \\
$n_e~(10^9~{\rm cm^{-3}})$ & $2.4-2.9$ , $2.8-3.3$ & $13-34$ & $0.06-24$ & $0.8-1.8$ & $5.0-26$ , $2.5-13$ & $\sim15$ & $0.8-17$ & $0.3-21$  \\
Radial size ($R_S$) & $44-52$ , $39-45$ & $2.2-5.6$ & $0.5-173$ & $35-81$ & $1.8-9.5$ , $0.6-2.9$ & $\sim0.5$ & $1.3-27$ & $2.8-210$  \\
\noalign{\smallskip}
\hline
\noalign{\smallskip}
\multicolumn{9}{c}{$f=3000$ (flat bread)} \\
\noalign{\smallskip}
\hline
\noalign{\smallskip}
$r$ (ld) & $2.3-2.5$, $3.1-3.3$ & $0.7-1.1$ & $0.2-3.6$ & $0.8-1.2$ & $0.05-0.1$, $0.5-1.2$ & $\sim0.2$ & $0.6-2.6$ & $0.03-0.3$  \\
$n_e~(10^{11}~{\rm cm^{-3}})$ & $2.3-2.8$, $2.7-3.2$ & $13-32$ & $0.06-23$ & $0.8-1.7$ & $4.8-25$, $2.4-12$ & $\sim15$ & $0.8-16$ & $0.3-20$ \\
Radial size ($R_S$) & $0.46-0.54$ , $0.40-0.47$ & $0.02-0.06$ & $0.005-1.8$ & $0.36-0.85$ & $0.02-0.10$ , $0.01-0.03$ & $\sim0.01$ & $0.01-0.3$ & $0.03-2.2$   \\
\noalign{\smallskip}
\hline
\end{tabular}
\end{sidewaystable*}

We also compare the distance of the obscure to those of BLR, dusty torus, and the innermost warm absorber from the literature. \citet{den09} measured a $3.8\pm0.8$ ld distance of the broad H$\beta$ line via reverberation mapping. This value is consistent with values tabulated in \citet{mar14}, which ranges from $\sim2$~ld (He {\sc i} $\lambda5876$) to $\sim6$~ld (Pa$\beta$ and Pa$\delta$). In Paper II, the estimated inner radius of the dusty torus is $\sim107$~ld following \citet{nen08}. We caution that the inner torus radius estimated with \citet{nen08} is valid for small dust grains. For large dust grains, one would expect a smaller distance value by a factor of $\sim4-5$ \citep[for NGC\,5548,][]{lan19}. The reverberation measurement of the inner radius of the dusty torus is $\sim20$~ld \citep{sug06}. In Paper II, the innermost warm absorber component has a distance of $\sim36-190$~ld. 

\begin{figure*} 
\centering
\includegraphics[width=.8\hsize, trim={0.5cm 11.2cm 1.5cm .0cm}, clip]{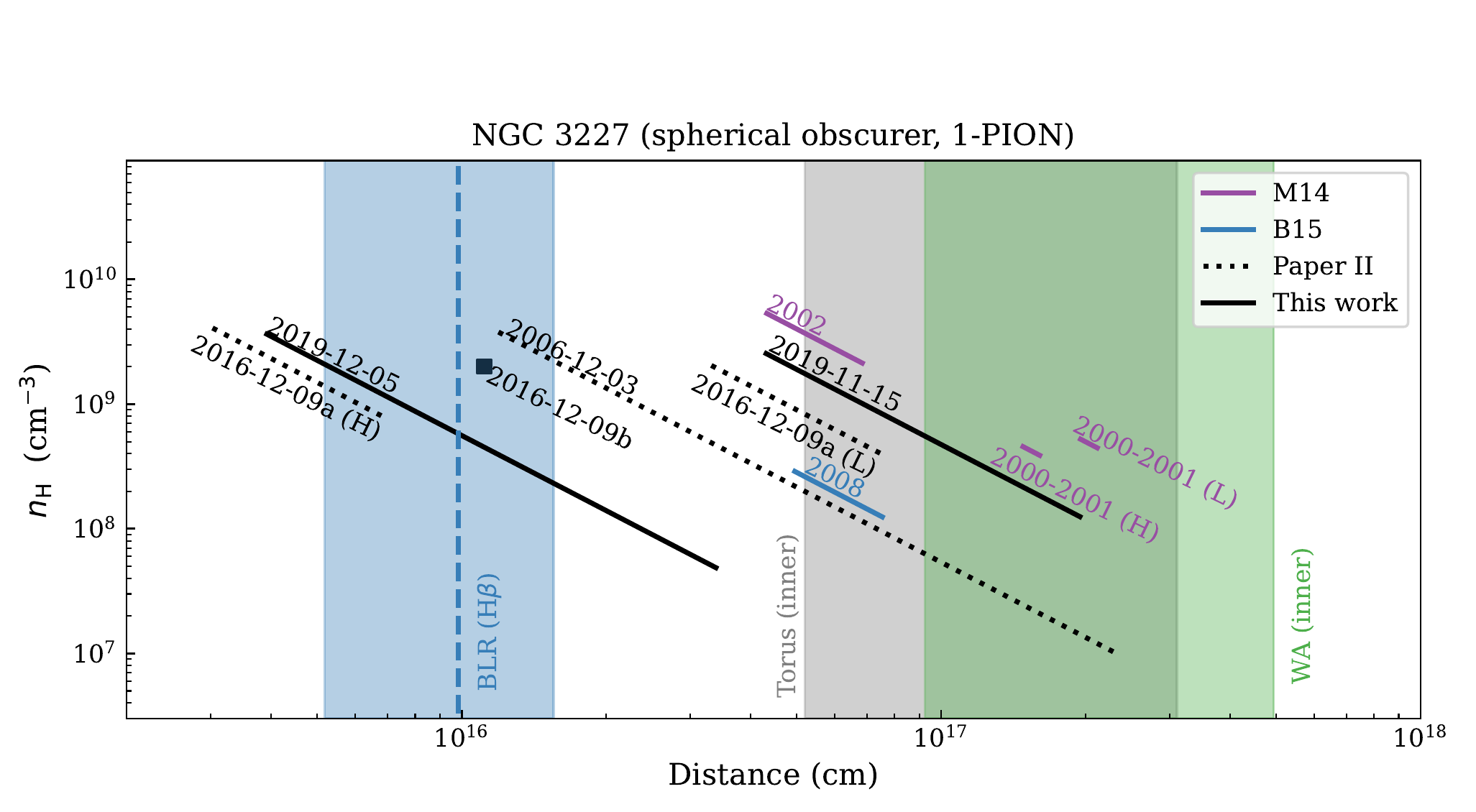}
\caption{Distance and number density ($n_{\rm H}$) of obscurers in NGC\,3227, assuming a spherical geometry ($f=1$ in Table~\ref{tbl:dist2BH}) for the obscurer. As the azimuthal to radial size ratio ($f$) increases, the estimated distance decreases while the density increases. M14 and B15 denotes obscuration events reported by \citet{mar14} and \citet{beu15}, respectively. Shaded area in blue mark the distance of the BLR from M14 ($2-6$~ld), as well as the H$\beta$ distance from \citet{den09}. Shaded area in grey mark the distance range of the inner radius of the dusty torus. The upper limit ($\sim107$~ld) is obtained for small dust grains (Eq.8 of Paper II). The lower limit \citep[$\sim20$~ld,][]{sug06} is the reverberation measurement. Shaded area in green mark the distance of the inner most warm absorber (Paper II). When multiple obscuration events occur on the same day, we use lower case letters (a and b) to differentiate them. When the obscurer requires two components with different ionization parameters, upper case letters H and L refer to the high- and low-ionization component, respectively. } 
\label{fig:plot_dist2BH}
\end{figure*}

Fig.~\ref{fig:plot_dist2BH} compares the distance of the spherical obscuring cloud and those of the BLR and dusty torus. As we can infer from Table~\ref{tbl:dist2BH}, as the azimuthal to radial size ratio ($f$) increases, the obscurer gets closer to the black hole ($r\propto f^{-2/5}$) and its number density increases while its radial size decreases. For $f\sim1$, the number density of the obscurer ($\sim10^{8-9}~{\rm cm^{-3}}$) is lower than the typical number density of the BLR clouds \citep[$\sim10^{9-13}~{\rm cm^{-3}}$,][]{pet06} but similar to the number density of torus materials \citep[$\sim10^8~{\rm cm^{-3}}$,][]{lan15}.

In our spectral analysis (Sect.~\ref{sct:spec}), we assume that all the warm absorber components are de-ionized by the obscurer. The same assumption is adopted in Papers I and II for NGC\,3227, as well as \citet{kaa14} for NGC\,5548 and \citet{meh17} for NGC\,3783. An obscurer with $r\lesssim107$~ld (the inner edge of the dusty torus) is consistent with our spectral analysis. This expectation can be met with $f\gtrsim1$ as shown in Table~\ref{tbl:dist2BH}. Nonetheless, we caution that it does not firmly rule out a geometry with $f\lesssim1$. In particularly, for $f\ll1$, our assumption of a uniform density and ionization parameter would likely fail due to stratification \citep[e.g.,][]{fuk14,mat20,gan21}. 

In Sect.~\ref{sct:spec}, we also fitted the observed spectra with two \textit{pion} components. For the 2019-11-15 observation, the highly ionized component of the obscurer only has an upper limit on the ionization parameter ($\log \xi \lesssim2$). For the 2019-12-05 observation, the lowly ionized component has very large uncertainties on the ionization parameter ($\log \xi = -0.5_{-3.8}^{+4.5}$). Thus, we cannot well constrain the distance of the obscuring components via Eq.~\ref{eq:dist2BH}.

In short, for an X-ray obscurer with one photoionized component, the effective UV emitting region radius ($R_{\rm 2500~\angstrom}\sim0.05$~ld, Sect.~\ref{sct:dist_uv}) is closer to the black hole than the obscurer unless the azimuthal to radial length scale of the obscurer is $\gg10^3$. Moreover, $R_{\rm 2500~\angstrom}\sim80~R_{\rm S}$ means that a spherical X-ray obscurer with its radius $\gtrsim80~R_{\rm S}$ is expected to intercept our line of sight to the UV emitting line region, whether the X-ray obscurer is close to the BLR or torus. Such deductions does not apply to other geometries with $f\ne1$. In Table~\ref{tbl:dist2BH}, for $f=1$, in only two out of the eight X-ray obscuration events (late 2000 to early 2001 and 2008), the radius (equals the radial size) of the X-ray obscurer is larger than $80~R_{\rm S}$. Unfortunately, we do not have UV grating spectroscopic observations in coordination with these X-ray observations.

\subsection{UV covering factor of the X-ray obscurer}
\label{sct:uv_fcov}
Since we cannot rule out the possibility that the obscurer is above the UV emitting region, here we explore possible interpretations of the observed data in this case.
The weakness or absence of the blueshifted broad absorption troughs in the HST/COS spectra can be explained if the obscurer observed in the X-ray band does not intercept our line of sight to (a significant portion of) the UV emitting region. Note that the X-ray covering factor of the obscurer is $\lesssim0.7$ for NGC\,3227 in 2019. The UV covering factor is expected to be even smaller. Assuming an fiducial X-ray emitting central engine of an order of $10~R_{\rm S}$, the UV emitting region with $R_{\rm 2500~\AA}\sim80~R_{\rm S}$ would be a factor of 64 larger. The X-ray obscurer would then cover $\lesssim1$~\% of the UV emitting region. Given the quality of our HST/COS spectra, it would be hard to detect broad absorption features with a few hundreds of km/s in width. In NGC\,5548, the X-ray covering factor between 2012 to 2016 is $\gtrsim0.7$ \citep{meh16b}. The relatively large inclination angle \citep[$\sim60^{\circ}$,][]{lyr13,fis13} for NGC\,3227 might also play a role. For comparison, the inclination angles for both NGC\,3783 and NGC\,5548 is $\sim40^{\circ}$ \citep[][]{lyr13,fis13,pan14}.

\section{Summary}
Multiple X-ray obscuration events have been reported in the nearby Seyfert 1.5 galaxy NGC\,3227 from 2000 to 2016. Another X-ray obscuration event was found in late 2019. Our photoionization modeling of the two X-ray observations (in mid-November and early December) requires distinct parameters of the obscurer, which cannot be explained by the same obscurer (whether it has one or two photoionized components) responding to the variable ionizing continuum.

In the UV band, previous X-ray obscuration events found in e.g., NGC\,5548 and NGC\,3783 are accompanied with blueshifted broad absorption troughs in the simultaneous UV grating spectra. However, no prominent blueshifted broad absorption troughs were found in NGC\,3227 when comparing the new HST/COS spectra obtained in 2019 with archival UV spectra obtained in 2000 and 2010.

We discuss two possible explanations for the lack of X-ray and UV association in NGC\,3227: (1) the obscurer is launched in the vicinity of the central engine and has not reached the UV emitting region yet; (2) the obscurer is above the UV emitting region but it is rather compact in size so that the X-ray obscurer does not cover a significant portion of the UV emitting region. For the former, due to the unknown geometry of the obscurer, we cannot well constrain its distance to the central engine and compared it to that of the UV emitting region. For the latter, we argued this might be the case based on the size of the X-ray and UV emitting region and the X-ray covering factor.

It is not straightforward to understand the variety of observational differences of the X-ray obscuration events in NGC\,3227 and other targets like NGC\,5548 and NGC\,3783. Future multi-wavelength spectroscopic observations are needed to establish a general understanding of the nature of the X-ray obscuration events in type I AGN.

\section*{Data availability}
A supplementary package is available at Zenodo \href{https://sandbox.zenodo.org/record/1039670#.Yjqlyn9BxjQ}{DOI: 10.5072/zenodo.1039670}. This package includes data and scripts used to reproduce the fitting results and figures presented in this work.

\begin{acknowledgements}
We thank the referee for careful reading and constructive suggestions to improve the quality of the manuscript. This work is based on observations obtained with XMM-Newton, an ESA science mission with instruments and contributions directly funded by ESA Member States and the USA (NASA). This research has made use of data obtained with the NuSTAR mission, a project led by the California Institute of Technology (Caltech), managed by the Jet Propulsion Laboratory (JPL) and funded by NASA. We thank the Swift team for monitoring our AGN sample, and the XMM-Newton, NuSTAR, and HST teams for scheduling our ToO triggered observations. This work work was supported by NASA through a grant for HST program number 15673 from the Space Telescope by the Association of Universities for Research in Astronomy, incorporated, under NASA contract NAS5-26555. SRON is supported financially by NWO, the Netherlands Organization for Scientific Research. JM acknowledges useful discussions with Zhu Liu. SGW acknowledges the support of a PhD studentship awarded by the UK Science \& Technology Facilities Council (STFC). SB acknowledges financial support from ASI under grants ASI-INAF I/037/12/0 and n. 2017-14-H.O and from PRIN MIUR project ``Black Hole winds and the Baryon Life Cycle of Galaxies: the stone-guest at the galaxy evolution supper", contract no. 2017PH3WAT. POP acknowledges financial support from the CNRS Programme National des Hautes Energies (PNHE) and from the french space agency CNES. BDM acknowledges support via Ram\'{o}n y Cajal Fellowship RYC2018-025950-I. DJW also acknowledges support from STFC in the form of an Ernest Rutherford Fellowship (ST/N004027/1). GP acknowledges funding from the European Research Council (ERC) under the European Union’s Horizon 2020 research and innovation programme (grant agreement No 865637). 
\end{acknowledgements}

%
%

\bibliographystyle{aa} 
\bibliography{refs} 

\begin{thebibliography}{71}
\expandafter\ifx\csname natexlab\endcsname\relax\def\natexlab#1{#1}\fi

\bibitem[{Antonucci(1993)}]{ant93}
Antonucci, R. 1993, ARA\&A, 31, 473

\bibitem[{{Arav} {et~al.}(2015){Arav}, {Chamberlain}, {Kriss}, {Kaastra},
  {Cappi}, {Mehdipour}, {Petrucci}, {Steenbrugge}, {et~al.}}]{ara15}
{Arav}, N., {Chamberlain}, C., {Kriss}, G.~A., {et~al.} 2015, \aap, 577, A37

\bibitem[{{Barret} {et~al.}(2018){Barret}, {Lam Trong}, {den Herder}, {Piro},
  {Cappi}, {Houvelin}, {Kelley}, {Mas-Hesse}, {Mitsuda}, {Paltani}, {Rauw},
  {Rozanska}, {Wilms}, {Bandler}, {Barbera}, {Barcons}, {Bozzo}, {Ceballos},
  {Charles}, {Costantini}, {Decourchelle}, {den Hartog}, {Duband}, {Duval},
  {Fiore}, {Gatti}, {Goldwurm}, {Jackson}, {Jonker}, {Kilbourne}, {Macculi},
  {Mendez}, {Molendi}, {Orleanski}, {Pajot}, {Pointecouteau}, {Porter},
  {Pratt}, {Pr{\^e}le}, {Ravera}, {Sato}, {Schaye}, {Shinozaki}, {Thibert},
  {Valenziano}, {Valette}, {Vink}, {Webb}, {Wise}, {Yamasaki}, {Douchin},
  {Mesnager}, {Pontet}, {Pradines}, {Branduardi-Raymont}, {Bulbul}, {Dadina},
  {Ettori}, {Finoguenov}, {Fukazawa}, {Janiuk}, {Kaastra}, {Mazzotta},
  {Miller}, {Miniutti}, {Naze}, {Nicastro}, {Scioritino}, {Simonescu},
  {Torrejon}, {Frezouls}, {Geoffray}, {Peille}, {Aicardi}, {Andr{\'e}},
  {Daniel}, {Cl{\'e}net}, {Etcheverry}, {Gloaguen}, {Hervet}, {Jolly}, {Ledot},
  {Paillet}, {Schmisser}, {Vella}, {Damery}, {Boyce}, {Dipirro}, {Lotti},
  {Schwander}, {Smith}, {Van Leeuwen}, {van Weers}, {Clerc}, {Cobo}, {Dauser},
  {Kirsch}, {Cucchetti}, {Eckart}, {Ferrando}, \& {Natalucci}}]{bar18}
{Barret}, D., {Lam Trong}, T., {den Herder}, J.-W., {et~al.} 2018, in Society
  of Photo-Optical Instrumentation Engineers (SPIE) Conference Series, Vol.
  10699, Space Telescopes and Instrumentation 2018: Ultraviolet to Gamma Ray,
  ed. J.-W.~A. {den Herder}, S.~{Nikzad}, \& K.~{Nakazawa}, 106991G

\bibitem[{{Bentz} \& {Katz}(2015)}]{ben15}
{Bentz}, M.~C. \& {Katz}, S. 2015, \pasp, 127, 67

\bibitem[{{Beuchert} {et~al.}(2015){Beuchert}, {Markowitz}, {Krau{\ss}},
  {Miniutti}, {Longinotti}, {Guainazzi}, {de La Calle P{\'e}rez}, {Malkan},
  {Elvis}, {Miyaji}, {Hiriart}, {L{\'o}pez}, {Agudo}, {Dauser}, {Garcia},
  {Kreikenbohm}, {Kadler}, \& {Wilms}}]{beu15}
{Beuchert}, T., {Markowitz}, A.~G., {Krau{\ss}}, F., {et~al.} 2015, \aap, 584,
  A82

\bibitem[{{Burke} {et~al.}(2021){Burke}, {Shen}, {Blaes}, {Gammie}, {Horne},
  {Jiang}, {Liu}, {McHardy}, {Morgan}, {Scaringi}, \& {Yang}}]{bur21}
{Burke}, C.~J., {Shen}, Y., {Blaes}, O., {et~al.} 2021, Science, 373, 789

\bibitem[{{Cappi} {et~al.}(2016){Cappi}, {De Marco}, {Ponti}, {Ursini},
  {Petrucci}, {Bianchi}, {Kaastra}, {Kriss}, {Mehdipour}, {Whewell}, {Arav},
  {Behar}, {Boissay}, {Branduardi-Raymont}, {Costantini}, {Ebrero}, {Di Gesu},
  {Harrison}, {Kaspi}, {Matt}, {Paltani}, {Peterson}, {Steenbrugge}, \&
  {Walton}}]{cap16}
{Cappi}, M., {De Marco}, B., {Ponti}, G., {et~al.} 2016, \aap, 592, A27

\bibitem[{{Crenshaw} {et~al.}(2003){Crenshaw}, {Kraemer}, \& {George}}]{cre03}
{Crenshaw}, D.~M., {Kraemer}, S.~B., \& {George}, I.~M. 2003, \araa, 41, 117

\bibitem[{{De Marco} {et~al.}(2020){De Marco}, {Adhikari}, {Ponti}, {Bianchi},
  {Kriss}, {Arav}, {Behar}, {Branduardi-Raymont}, {Cappi}, {Costantini},
  {Costanzo}, {di Gesu}, {Ebrero}, {Kaastra}, {Kaspi}, {Mao}, {Markowitz},
  {Matt}, {Mehdipour}, {Middei}, {Paltani}, {Petrucci}, {Pinto},
  {R{\'o}{\.z}a{\'n}ska}, \& {Walton}}]{dma20}
{De Marco}, B., {Adhikari}, T.~P., {Ponti}, G., {et~al.} 2020, \aap, 634, A65

\bibitem[{{De Rosa} {et~al.}(2015){De Rosa}, {Peterson}, {Ely}, {Kriss},
  {Crenshaw}, {Horne}, {Korista}, {Netzer}, {Pogge}, {Ar{\'e}valo}, {Barth},
  {Bentz}, {Brandt}, {Breeveld}, {Brewer}, {Dalla Bont{\`a}}, {De
  Lorenzo-C{\'a}ceres}, {Denney}, {Dietrich}, {Edelson}, {Evans}, {Fausnaugh},
  {Gehrels}, {Gelbord}, {Goad}, {Grier}, {Grupe}, {Hall}, {Kaastra}, {Kelly},
  {Kennea}, {Kochanek}, {Lira}, {Mathur}, {McHardy}, {Nousek}, {Pancoast},
  {Papadakis}, {Pei}, {Schimoia}, {Siegel}, {Starkey}, {Treu}, {Uttley},
  {Vaughan}, {Vestergaard}, {Villforth}, {Yan}, {Young}, \& {Zu}}]{dro15}
{De Rosa}, G., {Peterson}, B.~M., {Ely}, J., {et~al.} 2015, \apj, 806, 128

\bibitem[{{de Vaucouleurs} {et~al.}(1991){de Vaucouleurs}, {de Vaucouleurs},
  {Corwin}, {Buta}, {Paturel}, \& {Fouque}}]{dva91}
{de Vaucouleurs}, G., {de Vaucouleurs}, A., {Corwin}, Herold~G., J., {et~al.}
  1991, {Third Reference Catalogue of Bright Galaxies}

\bibitem[{{den Herder} {et~al.}(2001){den Herder}, {Brinkman}, {Kahn},
  {Branduardi-Raymont}, {Thomsen}, {Aarts}, {Audard}, {Bixler}, {den Boggende},
  {Cottam}, {Decker}, {Dubbeldam}, {Erd}, {Goulooze}, {G{\"u}del}, {Guttridge},
  {Hailey}, {Janabi}, {Kaastra}, {de Korte}, {van Leeuwen}, {Mauche},
  {McCalden}, {Mewe}, {Naber}, {Paerels}, {Peterson}, {Rasmussen}, {Rees},
  {Sakelliou}, {Sako}, {Spodek}, {Stern}, {Tamura}, {Tandy}, {de Vries},
  {Welch}, \& {Zehnder}}]{dhe01}
{den Herder}, J.~W., {Brinkman}, A.~C., {Kahn}, S.~M., {et~al.} 2001, \aap,
  365, L7

\bibitem[{{Denney} {et~al.}(2010){Denney}, {Peterson}, {Pogge}, {Adair},
  {Atlee}, {Au-Yong}, {Bentz}, {Bird}, {Brokofsky}, {Chisholm}, {Comins},
  {Dietrich}, {Doroshenko}, {Eastman}, {Efimov}, {Ewald}, {Ferbey}, {Gaskell},
  {Hedrick}, {Jackson}, {Klimanov}, {Klimek}, {Kruse}, {Lad{\'e}route}, {Lamb},
  {Leighly}, {Minezaki}, {Nazarov}, {Onken}, {Petersen}, {Peterson},
  {Poindexter}, {Sakata}, {Schlesinger}, {Sergeev}, {Skolski}, {Stieglitz},
  {Tobin}, {Unterborn}, {Vestergaard}, {Watkins}, {Watson}, \&
  {Yoshii}}]{den10}
{Denney}, K.~D., {Peterson}, B.~M., {Pogge}, R.~W., {et~al.} 2010, \apj, 721,
  715

\bibitem[{{Denney} {et~al.}(2009){Denney}, {Peterson}, {Pogge}, {Adair},
  {Atlee}, {Au-Yong}, {Bentz}, {Bird}, {Brokofsky}, {Chisholm}, {Comins},
  {Dietrich}, {Doroshenko}, {Eastman}, {Efimov}, {Ewald}, {Ferbey}, {Gaskell},
  {Hedrick}, {Jackson}, {Klimanov}, {Klimek}, {Kruse}, {Lad{\'e}route}, {Lamb},
  {Leighly}, {Minezaki}, {Nazarov}, {Onken}, {Petersen}, {Peterson},
  {Poindexter}, {Sakata}, {Schlesinger}, {Sergeev}, {Skolski}, {Stieglitz},
  {Tobin}, {Unterborn}, {Vestergaard}, {Watkins}, {Watson}, \&
  {Yoshii}}]{den09}
{Denney}, K.~D., {Peterson}, B.~M., {Pogge}, R.~W., {et~al.} 2009, \apjl, 704,
  L80

\bibitem[{{Di Gesu} {et~al.}(2015){Di Gesu}, {Costantini}, {Ebrero},
  {Mehdipour}, {Kaastra}, {Ursini}, {Petrucci}, {Cappi}, {Kriss}, {Bianchi},
  {Branduardi-Raymont}, {De Marco}, {De Rosa}, {Kaspi}, {Paltani}, {Pinto},
  {Ponti}, {Steenbrugge}, \& {Whewell}}]{dge15}
{Di Gesu}, L., {Costantini}, E., {Ebrero}, J., {et~al.} 2015, \aap, 579, A42

\bibitem[{{Ebrero} {et~al.}(2016){Ebrero}, {Kriss}, {Kaastra}, \&
  {Ely}}]{ebr16}
{Ebrero}, J., {Kriss}, G.~A., {Kaastra}, J.~S., \& {Ely}, J.~C. 2016, \aap,
  586, A72

\bibitem[{{Fischer} {et~al.}(2013){Fischer}, {Crenshaw}, {Kraemer}, \&
  {Schmitt}}]{fis13}
{Fischer}, T.~C., {Crenshaw}, D.~M., {Kraemer}, S.~B., \& {Schmitt}, H.~R.
  2013, \apjs, 209, 1

\bibitem[{{Fukumura} {et~al.}(2014){Fukumura}, {Tombesi}, {Kazanas}, {Shrader},
  {Behar}, \& {Contopoulos}}]{fuk14}
{Fukumura}, K., {Tombesi}, F., {Kazanas}, D., {et~al.} 2014, \apj, 780, 120

\bibitem[{{Gallo} {et~al.}(2021){Gallo}, {Gonzalez}, \& {Miller}}]{gal21}
{Gallo}, L.~C., {Gonzalez}, A.~G., \& {Miller}, J.~M. 2021, \apjl, 908, L33

\bibitem[{{Ganguly} {et~al.}(2021){Ganguly}, {Proga}, {Waters}, {Dannen},
  {Dyda}, {Giustini}, {Kallman}, {Raymond}, {Miller}, \& {Rodriguez
  Hidalgo}}]{gan21}
{Ganguly}, S., {Proga}, D., {Waters}, T., {et~al.} 2021, \apj, 914, 114

\bibitem[{{Gehrels} {et~al.}(2004){Gehrels}, {Chincarini}, {Giommi}, {Mason},
  {Nousek}, {Wells}, {White}, {Barthelmy}, {Burrows}, {Cominsky}, {Hurley},
  {Marshall}, {M{\'e}sz{\'a}ros}, {Roming}, {Angelini}, {Barbier}, {Belloni},
  {Campana}, {Caraveo}, {Chester}, {Citterio}, {Cline}, {Cropper}, {Cummings},
  {Dean}, {Feigelson}, {Fenimore}, {Frail}, {Fruchter}, {Garmire}, {Gendreau},
  {Ghisellini}, {Greiner}, {Hill}, {Hunsberger}, {Krimm}, {Kulkarni}, {Kumar},
  {Lebrun}, {Lloyd-Ronning}, {Markwardt}, {Mattson}, {Mushotzky}, {Norris},
  {Osborne}, {Paczynski}, {Palmer}, {Park}, {Parsons}, {Paul}, {Rees},
  {Reynolds}, {Rhoads}, {Sasseen}, {Schaefer}, {Short}, {Smale}, {Smith},
  {Stella}, {Tagliaferri}, {Takahashi}, {Tashiro}, {Townsley}, {Tueller},
  {Turner}, {Vietri}, {Voges}, {Ward}, {Willingale}, {Zerbi}, \&
  {Zhang}}]{geh04}
{Gehrels}, N., {Chincarini}, G., {Giommi}, P., {et~al.} 2004, \apj, 611, 1005

\bibitem[{{Grafton-Waters} {et~al.}(2020){Grafton-Waters},
  {Branduardi-Raymont}, {Mehdipour}, {Page}, {Behar}, {Kaastra}, {Arav},
  {Bianchi}, {Costantini}, {Ebrero}, {Di Gesu}, {Kaspi}, {Kriss}, {De Marco},
  {Mao}, {Middei}, {Peretz}, {Petrucci}, \& {Ponti}}]{gwa20}
{Grafton-Waters}, S., {Branduardi-Raymont}, G., {Mehdipour}, M., {et~al.} 2020,
  \aap, 633, A62

\bibitem[{{Green} {et~al.}(2012){Green}, {Froning}, {Osterman}, {Ebbets},
  {Heap}, {Leitherer}, {Linsky}, {Savage}, {Sembach}, {Shull}, {Siegmund},
  {Snow}, {Spencer}, {Stern}, {Stocke}, {Welsh}, {B{\'e}land}, {Burgh},
  {Danforth}, {France}, {Keeney}, {McPhate}, {Penton}, {Andrews},
  {Brownsberger}, {Morse}, \& {Wilkinson}}]{gre12}
{Green}, J.~C., {Froning}, C.~S., {Osterman}, S., {et~al.} 2012, \apj, 744, 60

\bibitem[{{Harrison} {et~al.}(2013){Harrison}, {Craig}, {Christensen},
  {Hailey}, {Zhang}, {Boggs}, {Stern}, {Cook}, {Forster}, {Giommi},
  {Grefenstette}, {Kim}, {Kitaguchi}, {Koglin}, {Madsen}, {Mao}, {Miyasaka},
  {Mori}, {Perri}, {Pivovaroff}, {Puccetti}, {Rana}, {Westergaard}, {Willis},
  {Zoglauer}, {An}, {Bachetti}, {Barri{\`e}re}, {Bellm}, {Bhalerao},
  {Brejnholt}, {Fuerst}, {Liebe}, {Markwardt}, {Nynka}, {Vogel}, {Walton},
  {Wik}, {Alexander}, {Cominsky}, {Hornschemeier}, {Hornstrup}, {Kaspi},
  {Madejski}, {Matt}, {Molendi}, {Smith}, {Tomsick}, {Ajello}, {Ballantyne},
  {Balokovi{\'c}}, {Barret}, {Bauer}, {Blandford}, {Brandt}, {Brenneman},
  {Chiang}, {Chakrabarty}, {Chenevez}, {Comastri}, {Dufour}, {Elvis}, {Fabian},
  {Farrah}, {Fryer}, {Gotthelf}, {Grindlay}, {Helfand}, {Krivonos}, {Meier},
  {Miller}, {Natalucci}, {Ogle}, {Ofek}, {Ptak}, {Reynolds}, {Rigby},
  {Tagliaferri}, {Thorsett}, {Treister}, \& {Urry}}]{har13}
{Harrison}, F.~A., {Craig}, W.~W., {Christensen}, F.~E., {et~al.} 2013, \apj,
  770, 103

\bibitem[{{Jansen} {et~al.}(2001){Jansen}, {Lumb}, {Altieri}, {Clavel}, {Ehle},
  {Erd}, {Gabriel}, {Guainazzi}, {Gondoin}, {Much}, {Munoz}, {Santos},
  {Schartel}, {Texier}, \& {Vacanti}}]{jan01}
{Jansen}, F., {Lumb}, D., {Altieri}, B., {et~al.} 2001, \aap, 365, L1

\bibitem[{{Kaastra}(2017)}]{kaa17}
{Kaastra}, J.~S. 2017, \aap, 605, A51

\bibitem[{{Kaastra} {et~al.}(2012){Kaastra}, {Detmers}, {Mehdipour}, {Arav},
  {Behar}, {Bianchi}, {Branduardi-Raymont}, {Cappi}, {Costantini}, {Ebrero},
  {Kriss}, {Paltani}, {Petrucci}, {Pinto}, {Ponti}, {Steenbrugge}, \& {de
  Vries}}]{kaa12}
{Kaastra}, J.~S., {Detmers}, R.~G., {Mehdipour}, M., {et~al.} 2012, \aap, 539,
  A117

\bibitem[{{Kaastra} {et~al.}(2014){Kaastra}, {Kriss}, {Cappi}, {Mehdipour},
  {Petrucci}, {Steenbrugge}, {Arav}, {Behar}, {Bianchi}, {Boissay},
  {Branduardi-Raymont}, {Chamberlain}, {Costantini}, {Ely}, {Ebrero}, {Di
  Gesu}, {Harrison}, {Kaspi}, {Malzac}, {De Marco}, {Matt}, {Nandra},
  {Paltani}, {Person}, {Peterson}, {Pinto}, {Ponti}, {Nu{\~n}ez}, {De Rosa},
  {Seta}, {Ursini}, {de Vries}, {Walton}, \& {Whewell}}]{kaa14}
{Kaastra}, J.~S., {Kriss}, G.~A., {Cappi}, M., {et~al.} 2014, Science, 345, 64

\bibitem[{{Kaastra} {et~al.}(2018){Kaastra}, {Mehdipour}, {Behar}, {Bianchi},
  {Branduardi-Raymont}, {Brenneman}, {Cappi}, {Costantini}, {De Marco}, {di
  Gesu}, {Ebrero}, {Kriss}, {Mao}, {Peretz}, {Petrucci}, {Ponti}, \&
  {Walton}}]{kaa18}
{Kaastra}, J.~S., {Mehdipour}, M., {Behar}, E., {et~al.} 2018, \aap, 619, A112

\bibitem[{{Kara} {et~al.}(2021){Kara}, {Mehdipour}, {Kriss}, {Cackett}, {Arav},
  {Barth}, {Byun}, {Brotherton}, {De Rosa}, {Gelbord}, {Hernandez Santisteban},
  {Hu}, {Kaastra}, {Landt}, {Li}, {Miller}, {Montano}, {Partington},
  {Aceituno}, {Bai}, {Bao}, {Bentz}, {Brink}, {Chelouche}, {Chen}, {Dalla
  Bonta}, {Dehghanian}, {Du}, {Edelson}, {Ferland}, {Ferrarese}, {Fian},
  {Filippenko}, {Fischer}, {Goad}, {Gonzalez Buitrago}, {Gorjian}, {Grier},
  {Guo}, {Hall}, {Homayouni}, {Horne}, {Ilic}, {Jiang}, {Joner}, {Kaspi},
  {Kochanek}, {Korista}, {Kynoch}, {Li}, {Liu}, {Mc Hardy}, {McLane},
  {Mitchell}, {Netzer}, {Olson}, {Pogge}, {Popovic}, {Proga},
  {Storchi-Bergmann}, {Strasburger}, {Treu}, {Vestergaard}, {Wang}, {Ward},
  {Waters}, {Williams}, {Yang}, {Yao}, {Zastrocky}, {Zhai}, \& {Zu}}]{kar21}
{Kara}, E., {Mehdipour}, M., {Kriss}, G.~A., {et~al.} 2021, arXiv e-prints,
  arXiv:2105.05840

\bibitem[{{Kriss} {et~al.}(2019{\natexlab{a}}){Kriss}, {De Rosa}, {Ely},
  {Peterson}, {Kaastra}, {Mehdipour}, {Ferland}, {Dehghanian}, {Mathur},
  {Edelson}, {Korista}, {Arav}, {Barth}, {Bentz}, {Brandt}, {Crenshaw}, {Dalla
  Bont{\`a}}, {Denney}, {Done}, {Eracleous}, {Fausnaugh}, {Gardner}, {Goad},
  {Grier}, {Horne}, {Kochanek}, {McHardy}, {Netzer}, {Pancoast}, {Pei},
  {Pogge}, {Proga}, {Silva}, {Tejos}, {Vestergaard}, {Adams}, {Anderson},
  {Ar{\'e}valo}, {Beatty}, {Behar}, {Bennert}, {Bianchi}, {Bigley}, {Bisogni},
  {Boissay-Malaquin}, {Borman}, {Bottorff}, {Breeveld}, {Brotherton}, {Brown},
  {Brown}, {Cackett}, {Canalizo}, {Cappi}, {Carini}, {Clubb}, {Comerford},
  {Coker}, {Corsini}, {Costantini}, {Croft}, {Croxall}, {Deason}, {De
  Lorenzo-C{\'a}ceres}, {De Marco}, {Dietrich}, {Di Gesu}, {Ebrero}, {Evans},
  {Filippenko}, {Flatland}, {Gates}, {Gehrels}, {Geier}, {Gelbord}, {Gonzalez},
  {Gorjian}, {Grupe}, {Gupta}, {Hall}, {Henderson}, {Hicks}, {Holmbeck},
  {Holoien}, {Hutchison}, {Im}, {Jensen}, {Johnson}, {Joner}, {Kaspi}, {Kelly},
  {Kelly}, {Kennea}, {Kim}, {Kim}, {Kim}, {King}, {Klimanov}, {Krongold},
  {Lau}, {Lee}, {Leonard}, {Li}, {Lira}, {Lochhaas}, {Ma}, {MacInnis},
  {Malkan}, {Manne-Nicholas}, {Matt}, {Mauerhan}, {McGurk}, {Montuori},
  {Morelli}, {Mosquera}, {Mudd}, {M{\"u}ller-S{\'a}nchez}, {Nazarov}, {Norris},
  {Nousek}, {Nguyen}, {Ochner}, {Okhmat}, {Paltani}, {Parks}, {Pinto},
  {Pizzella}, {Poleski}, {Ponti}, {Pott}, {Rafter}, {Rix}, {Runnoe}, {Saylor},
  {Schimoia}, {Schn{\"u}lle}, {Scott}, {Sergeev}, {Shappee}, {Shivvers},
  {Siegel}, {Simonian}, {Siviero}, {Skielboe}, {Somers}, {Spencer}, {Starkey},
  {Stevens}, {Sung}, {Tayar}, {Teems}, {Treu}, {Turner}, {Uttley}, {. Van
  Saders}, {Vican}, {Villforth}, {Villanueva}, {Walton}, {Waters}, {Weiss},
  {Woo}, {Yan}, {Yuk}, {Zheng}, {Zhu}, \& {Zu}}]{kri19b}
{Kriss}, G.~A., {De Rosa}, G., {Ely}, J., {et~al.} 2019{\natexlab{a}}, \apj,
  881, 153

\bibitem[{{Kriss} {et~al.}(2019{\natexlab{b}}){Kriss}, {Mehdipour}, {Kaastra},
  {Rau}, {Bodensteiner}, {Plesha}, {Arav}, {Behar}, {Bianchi},
  {Branduardi-Raymont}, {Cappi}, {Costantini}, {De Marco}, {Di Gesu}, {Ebrero},
  {Kaspi}, {Mao}, {Middei}, {Miller}, {Paltani}, {Peretz}, {Peterson},
  {Petrucci}, {Ponti}, {Ursini}, {Walton}, \& {Xu}}]{kri19a}
{Kriss}, G.~A., {Mehdipour}, M., {Kaastra}, J.~S., {et~al.} 2019{\natexlab{b}},
  \aap, 621, A12

\bibitem[{{Krolik} {et~al.}(1981){Krolik}, {McKee}, \& {Tarter}}]{kro81}
{Krolik}, J.~H., {McKee}, C.~F., \& {Tarter}, C.~B. 1981, \apj, 249, 422

\bibitem[{{Laha} {et~al.}(2021){Laha}, {Reynolds}, {Reeves}, {Kriss},
  {Guainazzi}, {Smith}, {Veilleux}, \& {Proga}}]{lah21}
{Laha}, S., {Reynolds}, C.~S., {Reeves}, J., {et~al.} 2021, Nature Astronomy,
  5, 13

\bibitem[{{Lamer} {et~al.}(2003){Lamer}, {Uttley}, \& {McHardy}}]{lam03}
{Lamer}, G., {Uttley}, P., \& {McHardy}, I.~M. 2003, \mnras, 342, L41

\bibitem[{{Landt} {et~al.}(2019){Landt}, {Ward}, {Kynoch}, {Packham},
  {Ferland}, {Lawrence}, {Pott}, {Esser}, {Horne}, {Starkey}, {Malhotra},
  {Fausnaugh}, {Peterson}, {Wilman}, {Riffel}, {Storchi-Bergmann}, {Barth},
  {Villforth}, \& {Winkler}}]{lan19}
{Landt}, H., {Ward}, M.~J., {Kynoch}, D., {et~al.} 2019, \mnras, 489, 1572

\bibitem[{{Landt} {et~al.}(2015){Landt}, {Ward}, {Steenbrugge}, \&
  {Ferland}}]{lan15}
{Landt}, H., {Ward}, M.~J., {Steenbrugge}, K.~C., \& {Ferland}, G.~J. 2015,
  \mnras, 449, 3795

\bibitem[{{Li} {et~al.}(2013){Li}, {Wang}, {Ho}, {Du}, \& {Bai}}]{lyr13}
{Li}, Y.-R., {Wang}, J.-M., {Ho}, L.~C., {Du}, P., \& {Bai}, J.-M. 2013, \apj,
  779, 110

\bibitem[{{Lodders} {et~al.}(2009){Lodders}, {Palme}, \& {Gail}}]{lod09}
{Lodders}, K., {Palme}, H., \& {Gail}, H.~P. 2009, Landolt B\&ouml;rnstein, 4B,
  712

\bibitem[{{Longinotti} {et~al.}(2019){Longinotti}, {Kriss}, {Krongold},
  {Arellano-Cordova}, {Komossa}, {Gallo}, {Grupe}, {Mathur}, {Parker},
  {Pradhan}, \& {Wilkins}}]{lon19}
{Longinotti}, A.~L., {Kriss}, G., {Krongold}, Y., {et~al.} 2019, \apj, 875, 150

\bibitem[{{Mao} {et~al.}(2018){Mao}, {Kaastra}, {Mehdipour}, {Gu},
  {Costantini}, {Kriss}, {Bianchi}, {Branduardi-Raymont}, {Behar}, {Di Gesu},
  {Ponti}, {Petrucci}, \& {Ebrero}}]{mao18}
{Mao}, J., {Kaastra}, J.~S., {Mehdipour}, M., {et~al.} 2018, \aap, 612, A18

\bibitem[{{Mao} {et~al.}(2017){Mao}, {Kaastra}, {Mehdipour}, {Raassen}, {Gu},
  \& {Miller}}]{mao17}
{Mao}, J., {Kaastra}, J.~S., {Mehdipour}, M., {et~al.} 2017, \aap, 607, A100

\bibitem[{{Mao} {et~al.}(2019){Mao}, {Mehdipour}, {Kaastra}, {Costantini},
  {Pinto}, {Branduardi-Raymont}, {Behar}, {Peretz}, {Bianchi}, {Kriss},
  {Ponti}, {De Marco}, {Petrucci}, {Di Gesu}, {Middei}, {Ebrero}, \&
  {Arav}}]{mao19}
{Mao}, J., {Mehdipour}, M., {Kaastra}, J.~S., {et~al.} 2019, \aap, 621, A99

\bibitem[{{Markowitz} {et~al.}(2014){Markowitz}, {Krumpe}, \&
  {Nikutta}}]{mar14}
{Markowitz}, A.~G., {Krumpe}, M., \& {Nikutta}, R. 2014, \mnras, 439, 1403

\bibitem[{{Matthews} {et~al.}(2020){Matthews}, {Knigge}, {Higginbottom},
  {Long}, {Sim}, {Mangham}, {Parkinson}, \& {Hewitt}}]{mat20}
{Matthews}, J.~H., {Knigge}, C., {Higginbottom}, N., {et~al.} 2020, \mnras,
  492, 5540

\bibitem[{{Mehdipour} {et~al.}(2016{\natexlab{a}}){Mehdipour}, {Kaastra}, \&
  {Kallman}}]{meh16a}
{Mehdipour}, M., {Kaastra}, J.~S., \& {Kallman}, T. 2016{\natexlab{a}}, \aap,
  596, A65

\bibitem[{{Mehdipour} {et~al.}(2017){Mehdipour}, {Kaastra}, {Kriss}, {Arav},
  {Behar}, {Bianchi}, {Branduardi-Raymont}, {Cappi}, {Costantini}, {Ebrero},
  {Di Gesu}, {Kaspi}, {Mao}, {De Marco}, {Matt}, {Paltani}, {Peretz},
  {Peterson}, {Petrucci}, {Pinto}, {Ponti}, {Ursini}, {de Vries}, \&
  {Walton}}]{meh17}
{Mehdipour}, M., {Kaastra}, J.~S., {Kriss}, G.~A., {et~al.} 2017, \aap, 607,
  A28

\bibitem[{{Mehdipour} {et~al.}(2016{\natexlab{b}}){Mehdipour}, {Kaastra},
  {Kriss}, {Cappi}, {Petrucci}, {De Marco}, {Ponti}, {Steenbrugge}, {Behar},
  {Bianchi}, {Branduardi-Raymont}, {Costantini}, {Ebrero}, {Di Gesu}, {Matt},
  {Paltani}, {Peterson}, {Ursini}, \& {Whewell}}]{meh16b}
{Mehdipour}, M., {Kaastra}, J.~S., {Kriss}, G.~A., {et~al.} 2016{\natexlab{b}},
  \aap, 588, A139

\bibitem[{{Mehdipour} {et~al.}(2021){Mehdipour}, {Kriss}, {Kaastra}, {Wang},
  {Mao}, {Costantini}, {Arav}, {Behar}, {Bianchi}, {Branduardi-Raymont},
  {Brotherton}, {Cappi}, {De Marco}, {Di Gesu}, {Ebrero}, {Grafton-Waters},
  {Kaspi}, {Matt}, {Paltani}, {Petrucci}, {Pinto}, {Ponti}, {Ursini}, \&
  {Walton}}]{meh21}
{Mehdipour}, M., {Kriss}, G.~A., {Kaastra}, J.~S., {et~al.} 2021, arXiv
  e-prints, arXiv:2106.14957

\bibitem[{{Miller} {et~al.}(2021){Miller}, {Zoghbi}, {Reynolds}, {Raymond},
  {Barret}, {Behar}, {Brandt}, {Brenneman}, {Draghis}, {Kammoun}, {Koss},
  {Lohfink}, \& {Stern}}]{mil21}
{Miller}, J.~M., {Zoghbi}, A., {Reynolds}, M.~T., {et~al.} 2021, \apjl, 911,
  L12

\bibitem[{{Murphy} {et~al.}(1996){Murphy}, {Lockman}, {Laor}, \&
  {Elvis}}]{mur96}
{Murphy}, E.~M., {Lockman}, F.~J., {Laor}, A., \& {Elvis}, M. 1996, \apjs, 105,
  369

\bibitem[{{Nenkova} {et~al.}(2008){Nenkova}, {Sirocky}, {Nikutta},
  {Ivezi{\'c}}, \& {Elitzur}}]{nen08}
{Nenkova}, M., {Sirocky}, M.~M., {Nikutta}, R., {Ivezi{\'c}}, {\v{Z}}., \&
  {Elitzur}, M. 2008, \apj, 685, 160

\bibitem[{{Netzer}(2015)}]{net15}
{Netzer}, H. 2015, \araa, 53, 365

\bibitem[{{Nicastro} {et~al.}(1999){Nicastro}, {Fiore}, {Perola}, \&
  {Elvis}}]{nic99}
{Nicastro}, F., {Fiore}, F., {Perola}, G.~C., \& {Elvis}, M. 1999, \apj, 512,
  184

\bibitem[{{Padovani} {et~al.}(2017){Padovani}, {Alexander}, {Assef}, {De
  Marco}, {Giommi}, {Hickox}, {Richards}, {Smol{\v{c}}i{\'c}},
  {Hatziminaoglou}, {Mainieri}, \& {Salvato}}]{pad17}
{Padovani}, P., {Alexander}, D.~M., {Assef}, R.~J., {et~al.} 2017, \aapr, 25, 2

\bibitem[{{Pancoast} {et~al.}(2014){Pancoast}, {Brewer}, {Treu}, {Park},
  {Barth}, {Bentz}, \& {Woo}}]{pan14}
{Pancoast}, A., {Brewer}, B.~J., {Treu}, T., {et~al.} 2014, \mnras, 445, 3073

\bibitem[{{Parker} {et~al.}(2019){Parker}, {Longinotti}, {Schartel}, {Grupe},
  {Komossa}, {Kriss}, {Fabian}, {Gallo}, {Harrison}, {Jiang}, {Kara},
  {Krongold}, {Matzeu}, {Pinto}, \& {Santos-Lle{\'o}}}]{par19}
{Parker}, M.~L., {Longinotti}, A.~L., {Schartel}, N., {et~al.} 2019, \mnras,
  490, 683

\bibitem[{{Peterson}(2006)}]{pet06}
{Peterson}, B.~M. 2006, {The Broad-Line Region in Active Galactic Nuclei}, ed.
  D.~{Alloin}, Vol. 693, 77

\bibitem[{{Risaliti} {et~al.}(2007){Risaliti}, {Elvis}, {Fabbiano}, {Baldi},
  {Zezas}, \& {Salvati}}]{ris07}
{Risaliti}, G., {Elvis}, M., {Fabbiano}, G., {et~al.} 2007, \apjl, 659, L111

\bibitem[{{Risaliti} {et~al.}(2011){Risaliti}, {Nardini}, {Salvati}, {Elvis},
  {Fabbiano}, {Maiolino}, {Pietrini}, \& {Torricelli-Ciamponi}}]{ris11}
{Risaliti}, G., {Nardini}, E., {Salvati}, M., {et~al.} 2011, \mnras, 410, 1027

\bibitem[{{Saez} {et~al.}(2021){Saez}, {Brandt}, {Bauer}, {Chartas}, {Misawa},
  {Hamann}, \& {Gallagher}}]{sae21}
{Saez}, C., {Brandt}, W.~N., {Bauer}, F.~E., {et~al.} 2021, \mnras, 506, 343

\bibitem[{{Serafinelli} {et~al.}(2021){Serafinelli}, {Braito}, {Severgnini},
  {Tombesi}, {Giani}, {Piconcelli}, {Della Ceca}, {Vagnetti}, {Gaspari},
  {Saturni}, {Middei}, \& {Tortosa}}]{ser21}
{Serafinelli}, R., {Braito}, V., {Severgnini}, P., {et~al.} 2021, arXiv
  e-prints, arXiv:2107.06584

\bibitem[{{Silva} {et~al.}(2016){Silva}, {Uttley}, \& {Costantini}}]{sil16}
{Silva}, C.~V., {Uttley}, P., \& {Costantini}, E. 2016, \aap, 596, A79

\bibitem[{{Suganuma} {et~al.}(2006){Suganuma}, {Yoshii}, {Kobayashi},
  {Minezaki}, {Enya}, {Tomita}, {Aoki}, {Koshida}, \& {Peterson}}]{sug06}
{Suganuma}, M., {Yoshii}, Y., {Kobayashi}, Y., {et~al.} 2006, \apj, 639, 46

\bibitem[{{Tarter} {et~al.}(1969){Tarter}, {Tucker}, \& {Salpeter}}]{tar69}
{Tarter}, C.~B., {Tucker}, W.~H., \& {Salpeter}, E.~E. 1969, \apj, 156, 943

\bibitem[{{Turner} {et~al.}(1996){Turner}, {George}, {Kallman}, {Yaqoob}, \&
  {Zycki}}]{tur96}
{Turner}, T.~J., {George}, I.~M., {Kallman}, T., {Yaqoob}, T., \& {Zycki},
  P.~T. 1996, \apj, 472, 571

\bibitem[{{Turner} {et~al.}(2018){Turner}, {Reeves}, {Braito}, {Lobban},
  {Kraemer}, \& {Miller}}]{tur18}
{Turner}, T.~J., {Reeves}, J.~N., {Braito}, V., {et~al.} 2018, \mnras, 481,
  2470

\bibitem[{{Wang} {et~al.}(2022){Wang}, {Kaastra}, {Mehdipour}, {Mao},
  {Costantini}, {Kriss}, {Pinto}, {Ponti}, {Behar}, {Bianchi},
  {Branduardi-Raymont}, {De Marco}, {Grafton-Waters}, {Petrucci}, {Ebrero},
  {Walton}, {Kaspi}, {Xue}, {Paltani}, {di Gesu}, \& {He}}]{wang22}
{Wang}, Y., {Kaastra}, J., {Mehdipour}, M., {et~al.} 2022, \aap, 657, A77

\bibitem[{{Whewell} {et~al.}(2015){Whewell}, {Branduardi-Raymont}, {Kaastra},
  {Mehdipour}, {Steenbrugge}, {Bianchi}, {Behar}, {Ebrero}, {Cappi},
  {Costantini}, {De Marco}, {Di Gesu}, {Kriss}, {Paltani}, {Peterson},
  {Petrucci}, {Pinto}, \& {Ponti}}]{whe15}
{Whewell}, M., {Branduardi-Raymont}, G., {Kaastra}, J.~S., {et~al.} 2015, \aap,
  581, A79

\bibitem[{{Wildy} {et~al.}(2021){Wildy}, {Landt}, {Ward}, {Czerny}, \&
  {Kynoch}}]{wil21}
{Wildy}, C., {Landt}, H., {Ward}, M.~J., {Czerny}, B., \& {Kynoch}, D. 2021,
  \mnras, 500, 2063

\bibitem[{{XRISM Science Team}(2020)}]{xri20}
{XRISM Science Team}. 2020, arXiv e-prints, arXiv:2003.04962

\end{thebibliography}



\end{document}